\documentclass[12pt]{iopart}

\usepackage[latin1]{inputenc}
\usepackage[english]{babel}
\usepackage{graphics}
\usepackage{hyperref}
\usepackage{url}
\usepackage{breakurl}

\usepackage{iopams}

\newcommand{\beq}{\begin{equation}}
\newcommand{\eeq}{\end{equation}}
\newcommand{\bea}{\begin{eqnarray}}
\newcommand{\eea}{\end{eqnarray}}

\newcommand{\HubbleComovil}{\mathcal{H}}

\newcommand{\bra}[1]{\langle {#1}|}
\newcommand{\ket}[1]{|{#1}\rangle}
\newcommand{\bbra}[1]{\Big\langle {#1}\Big|}
\newcommand{\bket}[1]{\Big|{#1}\Big\rangle}
\newcommand{\then}{\Rightarrow}

\newcommand{\dphi}{\delta\varphi}

\newcommand{\yRI}{\hat{y}^{(R,I)}}
\newcommand{\pyRI}{\hat{\pi}^{(R,I)}}

\newcommand{\expec}[1]{ \langle {#1} \rangle}

\newcommand{\tc}{\eta^c_k}
\newcommand{\tcn}[1]{\eta_k^{c_{{#1}}}}

\newcommand{\cre}{\hat{a}^{\dagger}} 

\newcommand{\ann}{\hat{a}} 

\newcommand{\dRI}{d^{(R,I)}}
\newcommand{\eRI}{e^{(R,I)}}
\newcommand{\cRI}{c^{(R,I)}}

\newsavebox{\eqnumref}
\newdimen\mathindent
\AtEndOfClass{\mathindent\leftmargini}

\newcommand{\mH}{\mathcal{H}}

\newcommand{\nk}{\textbf{k}}
\newcommand{\x}{\textbf{x}}
\newcommand{\y}{\textbf{y}}
\newcommand{\vac}{|0 \rangle}

\newcommand{\yk}{\hat{y}_{\textbf{k}}}
\newcommand{\pk}{\hat{\pi}_{\textbf{k}}}
\newcommand{\ak}{\hat{a}_{\textbf{k}}}
\newcommand{\aq}{\hat{a}_{\textbf{k}}}

\newcommand{\temp}{\frac{\delta T}{T_0}}
 \newcommand{\half}{\frac{1}{2}}
\newcommand{\mpl}{M_{P}}

\newcommand{\dyk}{\Delta\hat{y}_{\textbf{k}}}
\newcommand{\dpk}{\Delta\hat{\pi}_{\textbf{k}}}

\begin{document}

\title[Multiple quantum collapse of the inflaton field and the birth of cosmic structure]{Multiple quantum collapse of the inflaton field and its implications on the birth of cosmic structure}

\author{Gabriel León$^1$, Adolfo De Unánue$^2$  and Daniel Sudarsky$^3$\footnote{On sabbatical leave from: Instituto de Ciencias Nucleares, Universidad Nacional Autónoma de México, México D.F. 04510, México.  }}

\address{$^1$ Instituto de Ciencias Nucleares, Universidad Nacional Autónoma de México, México D.F. 04510, México}

\address{$^2$ C3 Centro de Ciencias de la Complejidad, Universidad Nacional Aut\'onoma de M\'exico, Torre de Ingenier\'ia, Circuito Exterior S/N Ciudad Universitaria, M\'exico D.F. 04510}

\address{$^3$ Instituto de Astronom\'ia y F\'isica del Espacio (UBA-CONICET), Casilla de Correos 67, Sucursal 28, 1428 Buenos Aires, Argentina}

\eads{\mailto{gabriel.leon@nucleares.unam.mx}, \mailto{adolfo@nucleares.unam.mx}, \mailto{sudarsky@nucleares.unam.mx}}

\begin{abstract}
  The standard inflationary account for the origin of cosmic structure
  is, without a doubt, extremely successful. However, it is not fully
  satisfactory as has been argued in [A. Perez, H. Sahlmann, and
  D. Sudarsky, Class. Quantum Grav., \textbf{23}, 2317, (2006)]. The
  central point is that, in the standard accounts, the inhomogeneity
  and anisotropy of our universe seems to emerge, unexplained, from an
  exactly homogeneous and isotropic initial state through processes
  that do not break those symmetries. The proposal made there to
  address this shortcoming calls for a dynamical and self-induced
  quantum collapse of the original homogeneous and isotropic state of
  the inflaton.  In this article, we consider the possibility of a
  multiplicity of collapses in each one of the modes of the Quantum
  Field. As we will see, the results are sensitive to a more detailed
  characterization of the collapse than those studied in the previous
  works, and in this regard two simple options will be studied. We
  find important constraints on the model, most remarkably on the
  number of possible collapses for each mode.
\end{abstract}


\submitto{\CQG}
\maketitle

\section{Introduction}

Modern cosmology has become a very successful field of research in
recent years. One of the major ideas, incorporated in the cosmological
model, is the existence of a period of accelerating expansion early in
the Universe's history, called Inflation. One of the major successes
of inflationary cosmology is its ability to `account for' the spectrum
of the temperature anisotropies in the Cosmic Microwave Background
(CMB), which is understood as the earliest observational data about
the primordial density fluctuations that seed the growth of structure
in our Universe.

However, when considering this account in more detail, one immediately
notes that there is something odd about it. Namely, that out of an
initial situation, which is taken to be perfectly isotropic and
homogeneous, and based on a dynamics that supposedly preserves those
symmetries, one ends with a non-homogeneous and non-isotropic
situation.

The problem described above, has been acknowledged by some
cosmologists\footnote{Sometimes this problem is formulated as the
  Quantum-to-Classical transition.} \cite{Padmanabhan96} and even by
some authors in recent textbooks
\cite{Mukhanov2005,Weinberg2008,Liddle2009}. Nevertheless, several
researchers in the field continue to hold the belief that the issues
have been successfully resolved
\cite{Kiefer98a,Kiefer98b,Kiefer00a,Kiefer08}. For an extensive
discussion about why the standard explanations doesn't solve this
problem, we invite  the reader to consult the reference
\cite{Sudarsky09}.

In a recent series of works
\cite{Sudarsky06a,Sudarsky06b,Sudarsky07a,Sudarsky07,Unanue2008,Sudarsky09,Unanue2010,Gabriel2010}
the problem has been analyzed leading to the conclusion that we need
some new physics to be able to fully address the problem. The
essential idea (as exposed in
\cite{Sudarsky06a,Sudarsky06b,Sudarsky07a,Sudarsky07,Unanue2008,Sudarsky09,Unanue2010,Gabriel2010})
is to introduce a new ingredient to the inflationary paradigm:
\emph{the self-induced collapse hypothesis:} a phenomenological model
incorporating the description of the effects of a dynamical collapse
of the wave function of the inflaton on the subsequent cosmological
evolution. The idea is inspired by L. Di\'osi
\cite{Diosi1984,Diosi,DiosiLajos07} and R. Penrose's arguments
\cite{Penrose96,Penrose94,Penrose02,Penrose05} in the sense that the
unification of quantum theory and the theory of gravitation would
likely involve modifications in both theories, rather than only the
latter as is more frequently assumed. Moreover, Penrose's idea is that
the resulting modifications of the former should involve something
akin to a self-induced collapse of the wave-function occurring when
the matter fields are in a quantum superposition corresponding to
space-time geometries which are `too different from each
other'. This sort of self-induced collapse would, in fact, be
occurring in rather common situations, and would ultimately resolve
the long standing `measurement problem' in quantum mechanics.

The collapse hypothesis in this context was originally inspired by
Penrose's ideas, however it might be compatible with other collapse
mechanisms which attempt to give a reasonable solution to the
measurement problem. In essence, the collapse hypothesis simply
sustains that something intrinsic to the system, i.e., independent of
observers, induces the collapse or reduction of the quantum mechanical
state of the system. Various proposals of that sort have been
considered
\cite{Ghirardi1986,Ghirardi1990,Bassi02,Bassi03,Bassi07-2},
and might
well be compatible with the self-induced collapse of the inflaton's
wave-function that we are considering. However, we are not following
any previous proposed scheme as the intention at this point is to
learn what characteristics are needed for it to work in the present
context. The point is that, in the case at hand, the collapse
hypothesis can be tested and exposed through strictly empirical
analyses.

The proposal is, at this stage of the analysis, a purely
phenomenological scheme. It does not attempt to
explain the process in terms of some specific new physical theory, but
merely give a rather general parametrization of the quantum transition
involved. We will refer to this phenomenological model as the
\emph{collapse scheme}. We will not further recapitulate the
motivations and discussion of the original proposal and instead refer
the reader to the above mentioned works.

Previous works along these lines have focused on the times of collapse
and the natural basis for the collapse \cite{Unanue2008}, and the
issue of fine-tuning of the inflaton potential in the collapse schemes
\cite{Gabriel2010}.  However, so far the analysis has been based on
the consideration of a single collapse of the inflaton's wave-function
for each mode. That limitation of scope has allowed the investigation
to proceed without the post-collapse state being characterized beyond
the specification of the expectation values of the field and the
conjugate momentum in the corresponding modes. The motivation of this
present paper is to extract more information about the collapse by
considering the possibility that \emph{multiple collapses } occur in
each mode, a consideration that requires a further specification of
the post-collapse states; in particular, we are going to focus in
models where the post-collapse states can be regarded as
\emph{coherent} or \emph{squeezed} states.

The article is organized as follows: In section \ref{sudarsky} we briefly
review the quantum mechanical treatment of the field's fluctuations
introducing the collapse hypothesis; we will emphasize how the
self-induced collapse proposal is contrasted with the observations
and, additionally, we will describe the three \emph{collapse schemes}
that have been studied so far, namely: Independent, Newtonian and
Wigner schemes. In section \ref{multiples} we will generalize the
collapse hypothesis of section \ref{sudarsky} to the case of multiple
collapses. In section \ref{caract}, we will characterize the multiple
post-collapse states and obtain new information about the parameters
describing the post-collapse state. Finally in section \ref{discusion} we
will end with a discussion of the results obtained in the previous
sections.

Regarding notation we will use signature $(- + + +)$ for the metric
and Wald's convention for the Riemann tensor. We will use units where
$c =1$ but will keep the gravitational constant $G$ and $\hbar$
explicit throughout the paper.

\section{\label{sudarsky}The collapse model for the quantum
  fluctuations in the inflationary scenario}

In this section we will review the formalism used in analyzing the
collapse process. The full formalism and motivation is presented in
\cite{Sudarsky06a,Sudarsky06b,Sudarsky07a,Sudarsky07}. We will use a
semi-classical description of gravitation in interaction with quantum
fields as reflected in the semi-classical Einstein's equation $G_{ab}
= 8 \pi G \langle \hat{T}_{ab} \rangle$, whereas the other fields are
treated in the standard quantum field theory (in curved space-time)
fashion. This is supposed to hold at all times except when a quantum
gravity induced collapse of the wave function occurs. At that point,
one would have to assume, that the excitation of the fundamental
quantum gravitational degrees of freedom must be taken into account,
with the corresponding breakdown of the semiclassical
approximation (the possible breakdown of the semi-classical
  approximation is formally represented by the presence of a term
  $Q_{ab}$ in the left hand side of the semi-classical Einstein's
  equation which is supposed to become nonzero only during the
  collapse of the quantum mechanical wave function of the matter
  fields, see \cite{Sudarsky06a} for the detailed discussion).

The starting point is the action of a scalar field minimally coupled
to gravity

\begin{equation}\label{actioncol} S[\phi] = \int d^4x
  \sqrt{-g} \bigg[ \frac{1}{16 \pi G} R[g_{ab}]
  -\half g^{ab} \nabla_a \phi \nabla_b \phi  - V[\phi] \bigg].
\end{equation}

One then splits the corresponding fields into their homogeneous part
and the perturbations. Thus the metric and the scalar fields are
written as $g= g_0 + \delta g$ and $\phi = \phi_0 + \dphi$.

With the appropriate choice of gauge\footnote{Although the equations
  in this gauge are formally identical to the gauge-independent
  equations \cite{Nakamura08}, the analysis done here requires the
  choosing of a specific gauge. One can not work with the so called
  `gauge invariant combinations', because in the approach followed
  here, the metric and field fluctuations are treated on a different
  footing. The metric is considered a classical variable (taken to be
  describing, in an effective manner, the deeper fundamental degrees
  of freedom of the quantum gravity theory that one envisions, lies
  underneath), while the matter fields, specifically the inflaton
  field perturbations are given a standard quantum field (in curved
  space-time) treatment, with the two connected trough the
  semiclassical Einstein's equations.  The choice of gauge implies
  that the time coordinate is attached to some specific slicing of the
  perturbed space-time, and thus, our identification of the
  corresponding hypersurfaces (those of constant time) as the ones
  associated with the occurrence of collapses,--something deemed as an
  actual physical change--, turns what is normally a simple choice of
  gauge into a choice of the distinguished hypersurfaces, tied to the
  putative physical process behind the collapse. This naturally leads
  to tensions with the expected general covariance of a fundamental
  theory, a problem that afflicts all known collapse models, and which
  in the non-gravitational settings becomes the issue of compatibility
  with Lorentz or Poincare invariance of the proposals.  We must
  acknowledge that this generic problem of collapse models is an open
  issue for the present approach. One would expect that its resolution
  would be tied to the uncovering of the actual physics behind what we
  treat here as the collapse of the wave function (which we view as a
  merely an effective description). As has been argued in related
  works, and in following ideas originally exposed by R. Penrose
  \cite{Penrose96,Penrose94,Penrose02,Penrose05}, we hold that the
  physics that lies behind all this, ties the quantum treatment of
  gravitation with the foundational issues afflicting quantum theory
  in general, and in particular those with connection to the
  `measurement problem'. } (we will work with the longitudinal gauge
also referred to as the Newtonian gauge) and ignoring the vector and
tensor part of the metric perturbations, the space-time metric can
then be described by the line element
\begin{equation}\label{metric}
  ds^2 = a(\eta)^2 [-(1+2\Psi(\eta,\x)) d\eta^2 +
  (1-2\Psi(\eta,\x)) \delta_{ij} dx^idx^j],
\end{equation}
where $\Psi(\eta,\x)$ is referred to as the \emph{Newtonian
  potential}.

The inflationary regime is characterized by a scale factor $a(\eta)
\approx -1/[H_I(1-\epsilon)\eta]$, with $H_I^2 \approx 8 \pi G V/3$
(which is Friedmann's equation) and $\epsilon \equiv \half
(M_{P}^2/\hbar) (\partial_\phi V/V)^2$ the slow-roll parameter (which
during inflation $\epsilon \ll 1$); $\mpl$ the reduced Planck mass
$M_{P}^2 \equiv \hbar/(8\pi G)$.

The normalization of the scale factor will be set so $a = 1$ at the
`present cosmological time'. The inflationary regime would end at
$\eta = \eta_r$, a value which is negative and very small in absolute
terms ($\eta_r \approx -10^{-22}$ Mpc). That is, the conformal time
$\eta$ during the inflationary era is in the range $-\infty < \eta <
\eta_r$, thus $\eta =0$ is a particular value of the conformal time
that does not correspond to the inflationary period, in fact, it
belongs to the radiation dominated epoch.

The background scalar field $\phi_0$ will be considered in the
slow-roll regime, i.e., $\phi_0' = -(a^3/3a')\partial_\phi V$, where
the primes denotes $\{\}' \equiv d/d\eta\{\}$.

Combining the background equations with Einstein's equations to first
order in the perturbations we obtain
\begin{equation}
  \nabla^2 \Psi + \mu \Psi = 4 \pi G (u \dphi +
  \phi_0' \dphi'),
\end{equation}
where $\mu \equiv \HubbleComovil^2-\HubbleComovil'$; $u \equiv
3\HubbleComovil \phi_0'+a^2\partial_\phi V[\phi]$ and $\HubbleComovil
\equiv a'(\eta)/a(\eta)$. If one uses the expressions for the scale
factor during a de Sitter phase then $\mu = 0$, while the slow-rolling
approximation $\phi_0' = -a^2 \partial_\phi V/3 \HubbleComovil$
corresponds to the condition $u=0$. Under those simplifying conditions
the last equation becomes a Poisson-like equation
\begin{equation}\label{eqpsi}
  \nabla^2 \Psi = 4\pi G \phi_0' \dphi'
  \equiv s \dphi',
\end{equation}
with $s \equiv 4\pi G \phi_0'$, which can be rewritten, by using the
slow-roll parameter, the background equation for $\phi_0'$ in the
slow-roll regime and Friedmann's equation, as $s\equiv a\hbar \sqrt{V
  \epsilon}/(\sqrt{6} \mpl^2)$.

The next step involves the quantization of the field fluctuation. We
emphasize that the background field $\phi_0$ is described in a
classical\footnote{By \emph{classical}, in this context, we mean that
  the homogeneous background field $\phi_0(\eta)$ is taken as an
  approximated description of the quantum quantity $\protect\langle
  \psi| \hat{\phi} (x,\eta) | \psi \protect\rangle$, where the state
  $|\psi \protect\rangle$ is the vacuum state of $\hat{\dphi}(x,\eta)$.
} fashion and it is only the fluctuation $\dphi$ which is subjected to
a quantum treatment.

Actually, it is convenient to work with the auxiliary field $y=a
\dphi$. The equation of motion for this field is

\begin{equation}\label{ymot} y''- \bigg( \nabla^2 + \frac{a''}{a} \bigg)
  y = 0 .\end{equation}

The conjugated canonical momentum of $y$ is $\pi = y'-ya'/a$. In order
to avoid infrared problems we will consider a restriction of the
system to a box of side L, with periodic boundary conditions. The
field and its momentum can be decomposed in Fourier's modes as

  \begin{equation}\label{campomomento}
    \hat{y}(\eta,\x) = \frac{1}{L^3}
    \sum_{\nk} e^{i\nk \cdot \x} \hat{y}_{\nk}(\eta), \qquad
    \hat{\pi}(\eta,\x) =
    \frac{1}{L^3} \sum_{\nk} e^{i\nk \cdot \x} \hat{\pi}_{\nk} (\eta) ,
  \end{equation}

with the wave vectors satisfying $k_i L = 2\pi n_i$ for $i=1,2,3$. The
field operator coefficients are further written as:
$\hat{y}_{\nk}(\eta) \equiv y_k (\eta) \hat{a}_{\nk} +
\overline{y}_{k} (\eta) \hat{a}_{-\nk}^\dag$ and $\hat{\pi}_{\nk}
(\eta) \equiv g_k (\eta) \hat{a}_{\nk} + \overline{g}_k (\eta)
\hat{a}_{-\nk}^\dag$. The functions $y_k(\eta)$ and $g_k(\eta)$
reflect the election of the vacuum state. In our case,
as is customarily done in the field, we choose the so called
Bunch-Davies vacuum \cite{Birrel94}, resulting from this choice

  \begin{equation} \label{eq:bunchdavies}
    y_k(\eta)=
    \frac{1}{\sqrt{2k}}\bigg(1 - \frac{i}{\eta k} \bigg) \exp
    (- ik\eta), \qquad
    g_k (\eta) = - i
    \sqrt{\frac{k}{2}} \exp (- ik\eta).
  \end{equation}

The vacuum state is defined by the condition $\hat{a}_{\nk} \vac = 0$
for all $\nk$, and can be easily seen to be homogeneous and isotropic
at all scales. The self collapse is assumed to operate in close
analogy with a `measurement' in the quantum-mechanical sense, but of
course, without any external apparatus or observer that could be
thought as performing the measurement. The self-induced collapse, is
assumed to occur, independently, for each mode of the field. That is,
one assumes that at a certain time $\eta_k^c$ (from now on we will
refer to this particular time as the \emph{time of collapse}) the
state of each mode $\nk$ of the field, which was initially the vacuum,
changes spontaneously into another state.  This self-collapse of the
wave-function is inspired by Penrose's ideas
\cite{Penrose96,Penrose94,Penrose02,Penrose05}, in which gravity plays
a fundamental role on the collapse of the wave-function and it does
not require outside observers who perform a measurement in order to
induce the collapse. The collapse scheme as employed here, however,
does not propose at this point a concrete physical mechanism behind
it, although one envisions that a more profound theory, presumably
derived from quantum gravity, will eventually account for it. These
ideas and motivations are discussed in great detail in
\cite{Sudarsky06a,Sudarsky06b,Sudarsky07a,Sudarsky07}. In order to
study the possibility of multiple collapses, we will see that more
detailed specifications of the states after the collapse are needed in
contrast with the works
\cite{Sudarsky06a,Sudarsky06b,Sudarsky07a,Sudarsky07}.

Following \cite{Sudarsky06a} it is convenient to decompose the field
$\hat{y}_{\nk}$ and its conjugated momentum $\hat{\pi}_{\nk}$ in their
real and imaginary parts which are completely Hermitian
$\hat{y}_{\nk}(\eta)=\hat{y}_{\nk}^R(\eta)+i\hat{y}_{\nk}^I(\eta)$ and
$\hat{\pi}_{\nk}(\eta)=
\hat{\pi}_{\nk}^R(\eta)+i\hat{\pi}_{\nk}^I(\eta)$ where

  \begin{equation}
   \yk^{(R,I)}(\eta) = \frac{1}{\sqrt{2}} \bigg(
    y_k(\eta)\ak^{(R,I)} + \overline{y}_k(\eta)\ak^{\dag (R,I)} \bigg),
    \eeq
    \beq
    \pk^{(R,I)}(\eta) = \frac{1}{\sqrt{2}} \bigg(
    g_k(\eta)\ak^{(R,I)} + \overline{g}_k(\eta)\ak^{\dag (R,I)} \bigg),
  \end{equation}

where
\begin{equation} \ak^R \equiv \frac{1}{\sqrt{2}}(\ak+\hat{a}_{-\nk}),
  \qquad \ak^I \equiv \frac{-i}{\sqrt{2}}(\ak-\hat{a}_{-\nk}).
\end{equation}

The commutators of the real and imaginary annihilation and creation
operators are
\beq\label{conmut}
\fl [\ak^R,\aq^{R \dag}] = \hbar L^3(\delta_{\nk,\nk'}+\delta_{\nk,-\nk'}), \qquad \qquad [\ak^I,\aq^{I \dag}]= \hbar L^3(\delta_{\nk,\nk'}-\delta_{\nk,-\nk'}).
\eeq

A full characterization of the state of each mode of the field would
require the specification all statistical moments. In previous works
\cite{Sudarsky06a,Unanue2008,Gabriel2010}, the collapse has been
characterized only in terms of the expectation values of field and of
the momentum conjugate for the new quantum state. However, in this
present work, as we are assuming the possibility of multiple
collapses, we will need to focus on the first two statistical moments:
the expectation value and the uncertainties (see section
\ref{multiples}).

For any state $\ket\Xi$ of the field $\hat{y}$, we introduce the
following quantities
\beq\label{eq:dce_def}
\fl d_{\nk}^{(R,I)} \equiv \expec{\ak^{(R,I)}}_\Xi, \qquad  c_{\nk}^{(R,I)} \equiv \expec{(\ak^{(R,I)})^2}_\Xi, \qquad e_{\nk}^{(R,I)} \equiv \expec{\ak^{(R,I) \dag}\ak^{(R,I)}}_\Xi.
\eeq

The expectation values of the field modes can be written as
\beq \label{eq:expectations}
\fl \expec{\yk^{(R,I)} (\eta)}_\Xi = \sqrt{2}\Re\left(y_k(\eta) d_{\nk}^{(R,I)}\right), \qquad \expec{\pk^{(R,I)} (\eta)}_\Xi = \sqrt{2} \Re\left(g_k(\eta) d_{\nk}^{(R,I)}\right),
\eeq

while their uncertainties are

    \begin{eqnarray}
      \label{eq:dy}
\fl      (\Delta \yk^{(R,I)} (\eta))^2_\Xi &=
      \Re\left(y_k^2(\eta) c_{\nk}^{(R,I)}\right)
       \nonumber \\
      &+ \frac{1}{2}|y_k(\eta)|^2\left(\hbar
        L^3+2e_{\nk}^{(R,I)}\right)
      -2\left[\Re\left(y_k(\eta)d_{\nk}^{(R,I)}\right)\right]^2,
    \end{eqnarray}
    \begin{eqnarray}
      \label{eq:dp}
\fl      (\Delta \pk^{(R,I)} (\eta))^2_\Xi &=
      \Re\left(g_k^2(\eta) c_{\nk}^{(R,I)}\right)
       \nonumber \\
      &+ \frac{1}{2}|g_k(\eta)|^2\left(\hbar
        L^3+2e_{\nk}^{(R,I)}\right)-2
      \left[\Re\left(g_k(\eta)d_{\nk}^{(R,I)}\right)\right]^2,
    \end{eqnarray}

specifically for the vacuum state $\vac$ one has, as expected,
$d_{\nk}^{(R,I)} = c_{\nk}^{(R,I)} = e_{\nk}^{(R,I)}= 0$, and thus $\expec{\yk^{(R,I)}(\eta)}_0=0$, $\expec{\pk^{(R,I)}(\eta)}_0 = 0$,
and their corresponding uncertainties
\beq\label{uvacio}
\fl    \left(\Delta \yk^{(R,I)}(\eta)\right)^2_0=\frac{1}{2}|y_k(\eta)|^2\hbar L^3, \qquad
    \left(\Delta \pk^{(R,I)}(\eta)\right)_0^2 =
    \frac{1}{2}|g_k(\eta)|^2\hbar L^3.
\eeq

Once we specify the expectation value of the field's modes
$\yRI_{\nk}$ and $\pyRI_{\nk}$ in the post collapse state $\ket{\Xi}$
at the time of collapse $\tc$ ($\ket{0} \to \ket{\Xi}$)

\beq\label{eq:collapsescheme}
\fl    \expec{\yRI_{\nk}(\eta_k^c)}_\Xi \equiv
    \langle \Xi | \yRI_{\nk}(\eta_k^c) | \Xi \rangle, \qquad
    \expec{\pyRI_{\nk}(\eta_k^c)}_\Xi \equiv \langle \Xi |
    \pyRI_{\nk}(\eta_k^c) | \Xi \rangle,
\eeq

we can obtain the expectation values evolved at any time after the
collapse, provided that there is no additional collapse. In fact, by
comparing \eref{eq:collapsescheme} with \eref{eq:expectations} we
obtain

\numparts
\beq\label{eq:evolucion-vepi-2}
\langle \hat\pi_{\nk}^{(R,I)} (\eta)\rangle_{\Xi} = A(\eta,\tc) \langle \hat\pi_{\nk}^{(R,I)} (\tc) \rangle_{\Xi}  +B(\eta,\tc) \langle \hat y_{\nk}^{(R,I)} (\tc) \rangle_{\Xi},
\eeq
\beq\label{eq:evolucion-vey-2}
\langle \hat y_{\nk}^{(R,I)}(\eta) \rangle_{\Xi} =C(\eta,\tc) \langle \hat\pi_{\nk}^{(R,I)} (\tc) \rangle_{\Xi} +D(\eta, \tc) \langle \hat y_{\nk}^{(R,I)} (\tc) \rangle_{\Xi},
\eeq
\endnumparts

where $A, B, C$ and $D$ are time dependent functions which describe
the temporal evolution of the quantum system between $\tc$ to
$\eta$. In particular, in the inflationary stage these functions are
\numparts
  \label{eq:operadores_evolucion}
  \begin{equation}
    A(\eta,\tc) = \cos (k\eta - k\tc) + \frac{\sin (k\eta - k\tc)}{k\tc} ,
  \end{equation}
  \begin{equation}
    B(\eta,\tc) = - k\sin (k\eta - k\tc),
  \end{equation}
  \beq
    C(\eta,\tc) = \frac{\cos (k\eta - k\tc)}{k}\left( \frac{1}{k\eta} - \frac{1}{k\tc}\right) +\frac{\sin (k\eta - k\tc)}{k}\left(\frac{1}{k^2 \eta \tc} + 1
    \right),
  \eeq
  \begin{equation}
    D(\eta, \tc) = \cos (k\eta - k\tc) - \frac{\sin (k\eta - k\tc)}{k\eta}.
  \end{equation}
\endnumparts

Equations \eref{eq:evolucion-vepi-2} and \eref{eq:evolucion-vey-2} can be rewritten in matrix form
\begin{equation}\label{matriz}
  \Upsilon (\eta,\Xi) = \textbf{U}(\eta,\tc) \Upsilon
  (\tc,\Xi),
\end{equation}
where

\beq
\Upsilon (\eta,\Xi) \equiv  \left( \begin{array}{c}
            \langle \pi_{\nk}^{(R,I)} (\eta)\rangle_{\Xi}       \\
            \langle y_{\nk}^{(R,I)}(\eta) \rangle_{\Xi}
           \end{array}
    \right),
\eeq

  \begin{equation}\textbf{U}(\eta,\tc) \equiv \left( \begin{array}{lr}
      A(\eta,\tc) & B(\eta,\tc) \\
      C(\eta,\tc) & D(\eta,\tc)
    \end{array}
    \right),
  \end{equation}

  \begin{equation} \Upsilon (\tc,\Xi) \equiv \left( \begin{array}{c}
      \langle \pi_{\nk}^{(R,I)} (\tc)\rangle_{\Xi} \\
      \langle y_{\nk}^{(R,I)}(\tc) \rangle_{\Xi}
    \end{array}
    \right).
  \end{equation}

In this notation, it is clear that the matrix $\textbf{U} (\eta,\tc)$
represents the standard unitary evolution (this refers to the
  standard quantum mechanical evolution of states or operators as it
  might be the case, and should not be taken to mean that the matrix
  $\textbf{U} (\eta,\tc)$ is unitary. It is not, and there is no
  reason for it to be so), for the expectation value of the fields,
from the time $\tc$ to the arbitrary time $\eta$.

The evolution of the uncertainties $(\Delta \yk^{(R,I)} (\eta))^2_\Xi$
and $(\Delta \pk^{(R,I)} (\eta))^2_\Xi$ depends on the specific
post-collapse state. In particular, the quantities $c_{\nk}^{(R,I)}$
and $e_{\nk}^{(R,I)}$ depend on the state after the collapse. That is,
once we specify the post-collapse state (and thus the quantities
$c_{\nk}^{(R,I)}$ and $e_{\nk}^{(R,I)}$ are fixed), we can use
\eref{eq:dy} and \eref{eq:dp} to obtain the evolution of the uncertainties.

\subsection{\label{observed}Connection to Observations}

In order to connect the predicted quantities with the observed ones,
we start from \eref{eqpsi}
\begin{equation*}
  \nabla^2 \Psi(\eta,\x) = s \dphi'(\eta,\x).
\end{equation*}

For the mode $\Psi_{\nk}$, after a Fourier's decomposition, we obtain
\begin{equation}\label{feqpsi}
  \Psi_{\nk} (\eta) = \frac{-s}{k^2}\dphi_{\nk}'(\eta).
\end{equation}

After describing the parametrization of the collapse in the previous
section, we proceed to evaluate the perturbed metric using the the
semi-classical Einstein's Field Equations: $G_{ab} = 8 \pi G
\expec{\hat{T}_{ab}}$ which we described at the beginning of this
section. To lowest order this set of equations reduces to
\begin{equation}
  \label{eq:semi-classical0}  
  \Psi_{\nk} (\eta) = \frac{-s}{a k^2}
  \expec{\pk(\eta)},
\end{equation}
where we used that $\expec{\hat{\dphi}'_{\nk}}_{\Xi}$ is connected to
the expectation value of the momentum field by
$\expec{\hat{\dphi}'_{\nk}}_{\Xi} = \expec{\pk}_{\Xi} / a(\eta)$ on
the state $\ket{\Xi}$.

Recalling that $s\equiv a\hbar \sqrt{V \epsilon}/\sqrt{6} \mpl^2$, the
expression for the Newtonian potential is

\begin{equation}
  \label{eq:semi-classical} \Psi_{\nk} (\eta) = -\frac{\hbar}{k^2
    \mpl^2} \sqrt{\frac{V \epsilon}{6}} \expec{\pk(\eta)}.
\end{equation}

We note that before the collapse occurs, the state of the field is the
Bunch-Davies vacuum for which $ \expec{\pk(\eta)}=0$, consequently
$\Psi_{\nk}(\eta)=0$, and the spacetime is homogeneous and isotropic
(at that scale).  However, after the collapse takes place, the new
state will generically have $ \expec{\pk(\eta)}\not=0$ and the
gravitational perturbations appear. That is, the onset of the
inhomogeneity and anisotropy at each scale is associated with the
first collapse of the corresponding mode.

In order to obtain a theoretical prediction and contrast it with the
observations, we strictly can not use the expression of
$\Psi_{\nk}(\eta)$ as given in \eref{eq:semi-classical} because it
was obtained using the slow-roll approximation which is only valid in
the inflationary epoch, while the observations made today by our satellites depend on the Newtonian potential at the last scattering surface. That is, the observations rely on $\Psi(\eta_D,\x_D)$, with $\eta_D$ the time of decoupling and
$\x_D = R_D (\sin \theta \sin \phi, \sin \theta \cos \phi, \cos \theta)$ where
$R_D$ is the radius of the last scattering surface, $\theta, \phi$ are
the standard spherical coordinates in the sky

The conformal time of decoupling
  lies in the matter dominated epoch. Nevertheless, we will work with
  the expression for $\Psi_{\nk} (\eta)$ in the radiation dominated era, extending if one wants its range of validity
  which is from $\eta_r$ to $\eta_{eq} < \eta_D$ (where $\eta_{eq}$ is
  the conformal time of the radiation-matter equality epoch). The
  changes during the brief period from the start of `matter
  domination' to `decoupling' (where the scale factor changes only by
  a factor of 3, i.e., $a(\eta_D)/a(\eta_{eq}) \approx 3$), are
  naturally considered to be irrelevant for the issues concerning us
  here, and thus the approximated value for the quantities of observational interest obtained
  using $\Psi (\eta)$ in the radiation dominated regime should be a very good approximation for the
  exact value of these quantities. Therefore, our goal here is to obtain an estimate for
$\Psi_{\nk}(\eta)$ during the radiation epoch.

The analysis can be
simplified by working with a quantity whose evolution is rather
simple, the so called `intrinsic curvature perturbation'
\cite{Bardeen83,Brandenberger84,Lyth85} $\zeta$, which is defined as
\begin{equation}
  \label{zeta} \zeta \equiv \frac{2}{3(w+1)} \left( \mH^{-1} \Psi' +
    \Psi \right) + \Psi,
\end{equation}

where $w\equiv P/\rho$. During the inflationary regime, $P= -\half
g^{ab} \partial_a \phi \partial_b \phi-V$ represents the `pressure' of
the scalar field and $\rho= -\half g^{ab} \partial_a \phi \partial_b
\phi+V$ the energy density.


It is a known result \cite{Mukhanov2005,Mukhanov90} that $\zeta$ is,
for modes larger than the Hubble radius (commonly referred as modes
`larger than the horizon' i.e., modes with $k \ll \mH$) and for `adiabatic
perturbations', roughly a `constant quantity', irrespective of the
cosmological regime and the nature of the dominant kind of matter. The
constancy of this quantity is used to obtain a relation between the
values of the Newtonian potential during the two relevant regimes:
$\Psi^{inf}_{\nk}(\eta)$ and $\Psi^{rad}_{\nk}(\eta)$

\beq\label{zetab}
  \zeta^{inf} = \zeta^{rad} \qquad \then
  \qquad
  \Psi^{inf}_{\nk} \bigg[ \frac{2}{3} \bigg( \frac{1}{w_{inf} + 1}
  \bigg) +1\bigg] = \frac{3}{2} \Psi^{rad}_{\nk},
\eeq
where, in obtaining the right hand side of \eref{zetab} the use of
the equation of state $P = \rho/3$ was made, and the left hand side
was obtained using the equation of state $P = w_{inf} \rho$ where
$w_{inf} + 1 = \phi_0'^2/ a^2
\rho$. 
Finally, by relying on the assumption of validity of the slow-roll
approximation during inflation, $\phi_0'^2/a^2 = \frac{2}{3} V
\epsilon$, \eref{zetab} becomes
\begin{equation}\label{psiinfrad0} \Psi^{rad}_{\nk} = \frac{2}{3}
  \frac{\Psi^{inf}_{\nk}}{\epsilon}.  \end{equation}
Thus,  substituting
\eref{eq:semi-classical} in \eref{psiinfrad0}, the expression for
the Newtonian potential, in the radiation dominated epoch, becomes
\begin{equation}\label{psiinfrad}
  \Psi^{rad}_{\nk} (\eta) = \frac{-\hbar}{ 2 \mpl^2}
  \sqrt{ \frac{8V}{27\epsilon} } \frac{\expec{\pk(\eta)}}{k^2}.
\end{equation}

The expression above is valid for modes with $k \ll \mH$, which are
actually the modes of interest from the observational point of view.
That is, we need to consider that $k/\mH \ll 1$ in
$\expec{\pk(\eta)}$. Furthermore, the result \eref{psiinfrad} shows
that for a generic collapse scheme there is an amplification
$1/\epsilon$ in the Newtonian potential, in accordance with the
generic findings of the detailed study for the collapse scheme
presented in \cite{Gabriel2010}.

In order to connect with the observations we note that the quantity
that is observed is $\temp (\theta, \phi)$, which is expressed in
terms of its spherical harmonic decomposition $\sum_{lm} \alpha_{lm}
Y_{lm} (\theta , \phi)$. The theoretical calculations make a
prediction for the most likely value of the coefficients $\alpha_{lm}$
which are expressed in terms of the Newtonian potential on the
2-sphere corresponding to the intersection of our past light cone with
the surface of last scattering

\begin{equation}
  \alpha_{lm} = \int d^2 \Omega \Psi (\eta_D,\x_D) Y_{lm} (\theta, \phi),
\end{equation}

After a Fourier decomposition of the
Newtonian potential $\Psi(\eta_D,\x_D) = \sum_\nk (\Psi_\nk
(\eta_D)/L^3) e^{i \nk \cdot \x_D}$, and using \eref{psiinfrad}, we
obtain

\begin{equation}
  \alpha_{lm} = \int d^2 \Omega \sum_{\nk}
  \frac{-\hbar}{ 2 \mpl^2 k^2 L^3} \sqrt{ \frac{8V}{27\epsilon} }
  \expec{\pk (\eta_D)} Y^\star_{lm}
  (\theta,\phi) e^{i \nk \cdot \x_D}.
\end{equation}

Using standard spherical harmonic relations: \\
$e^{i \nk \cdot \x_D} = 4
\pi \sum_{lm} i^l j_l(k R_D) Y_{lm} (\theta, \phi) Y^\star_{lm}
(\hat{\nk})$, where $j_l$ are spherical Bessel functions, we get

\begin{equation}
  \alpha_{lm} = 4\pi i^l \sum_{\nk} \frac{-\hbar}{ 2  \mpl^2 k^2 L^3}
  \sqrt{ \frac{8V}{27\epsilon} }  \expec{\pk (\eta_D)} j_l(k R_D)
  Y^\star_{lm} (\hat{\nk}).
\end{equation}

The quantity $\alpha_{lm}$ is the sum of contributions
from the collection of modes, each contribution being a complex
number, leading to what is in effect a sort of `two-dimensional random
walk' whose total displacement corresponds to the observational
quantity (this will be seen more clearly in the next section when we specify  $\expec{\pk (\tc)}$) .  It is clear that, as in the case of any random walk, such
quantity can not be evaluated and the only thing that can be done is
to evaluate the most likely value for such total displacement, with
the expectation that the observed quantity will be close to that
value. As is now standard in our treatments, we do this with the help
of the imaginary ensemble of universes and the identification of the
most likely value with the ensemble's mean value

  \begin{equation}\label{alphalm}
\fl    |\alpha_{lm}|^2_{M.L.} = \frac{(4 \pi)^2}{L^6}
    \sum_{\nk \nk'}   \frac{2\hbar^2 V}{27\epsilon \mpl^4}  \frac{1}{
      k^2 k'^2} \overline{\langle \pk (\eta_D)
      \rangle \langle \hat{\pi}^\dag_{\nk'} (\eta_D)\rangle} j_l (k R_D) j_l
    (k' R_D) Y^\star_{lm} (\hat{\nk}) Y_{lm} (\hat{\nk}').
  \end{equation}

The rest of the present work focuses on obtaining the quantity
$\overline{\langle \pk (\eta) \rangle \langle \hat{\pi}^\dag_{\nk'}
  (\eta) \rangle}$ under specific conditions on the post-collapses
states.

An important observation follows directly from the point of view
adopted to relate the metric effective description of gravity with the
quantum aspect of the matter fields: The source of the fluctuations
that lead to anisotropies and inhomogeneities lies in the quantum
uncertainties for the scalar field, which collapses, due to some
unknown quantum gravitational effect. Once collapsed, these density
inhomogeneities and anisotropies feed into the gravitational degrees
of freedom leading to nontrivial perturbations in the metric
functions, in particular, the Newtonian potential. However, the metric
itself is not a source of the quantum gravitational induced
collapse. Therefore, as the scalar field does not act as a source for
the gravitational tensor modes -at least not at the lowest order
considered here-, the tensor modes can not be excited. Thus, as
already discussed in \cite{Sudarsky06a,Sudarsky06b}, the scheme
naturally leads to the prediction\footnote{However, it is  worthwhile   pointing   out that  such conclusion   is  directly tied to
our  underlying approach that favours  the  semi-classical Einstein's  equations
augmented  with a collapse  proposal as  a way to deal with the gravity quantum interface
 faced in the  current problem. It is of course  conceivable, although seems harder to
understand in a wider context (see  the  discussion  in  section  8  of \cite{Sudarsky09}),  that a
collapse  might be incorporated into a setting where both  the gravitation and scalar
filed perturbations  are  simultaneously  treated  at the quantum  level. If the latter
happened to be the correct approach, something that would be possible to ascertain when
we have a fully satisfactory theory of quantum gravity, our conclusion about the tensor
modes would be modified.}  of a zero -or at least a strongly
suppressed- amplitude of gravitational waves to the CMB.

\subsection{Quantum collapse schemes}

In order to proceed, we must specify the quantum collapse scheme which
drives the inflaton field out of homogeneity and isotropy. In past
works \cite{Sudarsky06a,Unanue2008} three different schemes were
considered. Two of them, called \textit{Independent} collapse and
\textit{Newtonian} collapse were presented in \cite{Sudarsky06a} and
the last one, denominated \textit{Wigner's} collapse, was presented in
\cite{Unanue2008}.  In \cite{Unanue2008} these schemes are further
studied but limiting the consideration to a single collapse. In the
following, we will describe them briefly.

\subsubsection{\label{eq:esquema_independiente}Independent collapse
  scheme}

In this scheme one assumes that the expectation values of the field's
mode $\yRI_{\nk}$, and their conjugate momentum $\pyRI_{\nk}$, acquire
`random' independent values. The expectation value was considered as
randomly selected

\beq
\fl  \expec{\yRI_{\nk}(\eta^c_{\nk})}_\Xi = x_{\nk,I}^{(R,I)}
  \sqrt{\left(\Delta\yk^{(R,I)} (\eta_k^c) \right)^2_0}, \qquad
  \expec{\pyRI_{\nk}(\eta^c_k)}_\Xi = x_{\nk,II}^{(R,I)}
  \sqrt{\left(\Delta\pyRI_{\nk} (\eta_k^c) \right)^2_0}.
\end{equation}

In this scheme the expectation value jumps to a random value
$x_{\nk}^{(R,I)}$ multiplied by the uncertainty of the vacuum state of
the field. The random variables $x_{\nk,I}^{(R,I)}$,
$x_{\nk,II}^{(R,I)}$ are selected from a Gaussian distribution
centered at zero, of spread one (normalized), and are statistically
uncorrelated, that is the rationale of the name. This means that we
are ignoring the \textit{natural} correlation that exists in the
conjugate fields in the pre-collapse state.

\subsubsection{Newtonian collapse scheme}

This scheme is motivated by the observation that in the Poisson-like
equation \eref{eq:semi-classical}, only the expectation value of
$\pyRI$ appears. Thus, following Penrose's ideas regarding the quantum
uncertainties that the gravitational potential would be inheriting
from the matter fields' quantum uncertainties, as fundamental factors
triggering the collapse, one is led to consider a scheme where `only
$\pyRI$ collapses', leaving the expectation value of $\yRI$ unchanged
\begin{equation}
  \expec{\yRI_{\nk}(\eta^c_k)}_\Xi = 0, \qquad
  \expec{\pyRI_{\nk}(\eta^c_k)}_\Xi = x_{\nk,II}^{(R,I)}
  \sqrt{\left(\Delta\pyRI_{\nk} (\eta_k^c) \right)^2_0}.
\end{equation}
As before, $x_{\nk,II}^{(R,I)}$ represents a random Gaussian variable
normalized and centered at zero.

\subsubsection{Wigner's collapse scheme}

The last collapse scheme considered in \cite{Unanue2008, Unanue2010}
attempts to take into account the correlation between $\yRI$ and
$\pyRI$ existing in the pre-collapse state, and to characterize it in
terms of the Wigner's function. The Wigner's function of the vacuum
state of the inflaton is a bi-dimensional Gaussian function. This fact
will be used to model the resulting collapse of the quantum field
state. The assumption will be that, at a certain (conformal) time
$\eta^c_k$, the part of the state characterizing the mode $k$ will
collapse, leading to a new state in which the fields will have
expectation values given by
\begin{equation}
  \expec{\yRI_{\nk} (\eta_k^c) }_\Xi  = x^{(R,I)}_{\nk} \Lambda_k
  \cos\Theta_k, \qquad
  \expec{\pyRI_{\nk} (\eta_k^c) }_\Xi = x^{(R,I)}_{\nk}
  \Lambda_k k
  \sin\Theta_k,
\end{equation}
where $x_{\nk}^{(R,I)}$ is a random variable, characterized by a
Gaussian distribution centered at zero with a spread one. $\Lambda_k$
is given by the major semi-axis of the ellipse characterizing the
bi-dimensional Gaussian function (the ellipse corresponds to the
boundary of the region in `phase space' where the Wigner function has
a magnitude larger than 1/2 its maximum value), and $\Theta_k$ is the
angle between that axis and the $\hat{y}_{\nk}^{(R,I)}$ axis. The
quantities $\Lambda_k$ and $\Theta_k$ can be expressed in terms of
$\eta^c_k$ \cite{Unanue2008} as
\begin{equation}
  \Lambda_k   =  \frac{4\tc\sqrt{\hbar L^3
      k}}{\sqrt{1+5(k\tc)^2
      -\sqrt{1+10(k\tc)^2+9(k\tc)^4}}},
\end{equation}
\begin{equation}
  2  \Theta_k   =
  \arctan\left(\frac{4k\tc}{1-3(k\tc)^2}\right).
\end{equation}

\section{\label{multiples}Multiple Quantum Collapses}

Once one hypothesizes that there is a new kind of physical process
which affects the system under investigation, it seems logical to
consider the possibility that it occurs more than once, and in
circumstances different from those for which it was first proposed.
The extensive study of that issue is well beyond the present
manuscript and would require
its merger with other studies of collapse models for more general
circumstances. However, the cosmological situation is one where
further analysis can be done with relative ease and where it is
natural to assume that the effect must manifest in a rather unmodified
version in all its occurrences. This work can be considered as the
first exploration (see also \cite{Unanue2010}) of the effects of
multiple collapses in the situation which lead to the first proposals
suggesting that they play a fundamental role in cosmology.

\subsection{Collapse Scheme Relations for Multiple Collapses}

We will analyze the provided schemes in the case of multiple quantum
collapses by focusing on the expectation values of the relevant
quantities at any time after exactly $n$ quantum collapses.

First, we will generalize the notation in order to handle in a unified
fashion the three different quantum collapse schemes (developed in
\cite{Sudarsky06a,Unanue2008}). Let us assume that at time
$\eta_k^{c_1}$ has occurred a single collapse taking the state $|c_0
\rangle$ to the state $\ket{c_1}$\footnote{In this section, we have
  changed the notation slightly, the post-collapse state will be
  denoted $|c_n\protect\rangle$ instead of $|\Xi\protect\rangle$ as in
  the previous section. That is, the state $|c_o\protect\rangle$,
  represents the vacuum state; $|c_1\protect\rangle$ will denote the
  first collapse state and so on.}. Then, a natural generalization of
the expectation value of the field (and its conjugated momentum) in
the post-collapse state $\ket{c_1}$ will be assumed to be given by

  \beq\label{esquemagen1}
    \expec{\yRI_{\nk} (\eta_k^{c_1})}_{c_1}
    =  x_{\nk, I}^{(1) (R,I)} \sigma_{y}^{(R,I)}( \eta_k^{c_1}, c_0)
    +\expec{\yRI_{\nk} (\eta_k^{c_1})}_{c_0},
  \eeq
  \beq\label{esquemagen2}
    \expec{\pyRI_{\nk}(\eta_k^{c_1})}_{c_1} = x_{\nk, II}^{(1) (R,I)}
    \sigma_{\pi}^{(R,I)}( \eta_k^{c_1},
    c_0)
    +\expec{\pyRI_{\nk}(\eta_k^{c_1})}_{c_0},
  \eeq

where $x_{\nk, I}^{(1) (R,I)} $ and $x_{\nk, II}^{(1) (R,I)}$ stand
for a random value characterizing the change in the expectation value
of $\yRI_{\nk}$ and $\pyRI_{\nk}$ respectively. The superscript
$^{(1)}$ indicates that the random variables are associated with the
first collapse while the quantities in the last term of the right hand
side of \eref{esquemagen1} and \eref{esquemagen2} represent the value of
the corresponding operators if there had been no collapse. The
correlation between these random variables depends on the particular
collapse scheme.  The functions $\sigma_{y}^{(R,I)}(
\eta_k^{c_1},c_0)$ and $\sigma_{\pi}^{(R,I)}( \eta_k^{c_1},c_0)$
denote the uncertainties of the expectation values of the fields for
the particular collapse scheme considered. The notation employed
remind us the principal quantities that characterize the expectation
values.  That is, it depends on: the previous collapse state
$\ket{c_0}$ (which in the case of a single collapse is the vacuum
state), the time of collapse $\eta_k^{c_1}$, and the random variables
$x_{\nk, I}^{(1) (R,I)}, x_{\nk, II}^{(1) (R,I)}$.

Note that the left hand side (l.h.s.) of
\eref{esquemagen1} and \eref{esquemagen2} is in the post-collapse state
$\ket{c_1}$, while the right hand side (r.h.s.) is in the pre-collapse
state $\ket{c_0}$, i.e.,\ the new state depends on the old state. Let
us also note that the whole expressions
\eref{esquemagen1} and \eref{esquemagen2} are evaluated in $\tcn{1}$, the
time at which the first collapse occurs. The second term in the
r.h.s. is the expectation value of the mode in the state $|c_0
\rangle$ \textit{evolved} up to $\tcn{1}$. This dynamical evolution is
dictated by \eref{matriz}. The generalization of these ideas allow us
to write the collapse scheme for the $n$-th collapse $\ket{c_{n-1}}
\to \ket{c_n}$

  \beq\label{esquemagenn1}
    \langle \yRI_{\nk} (\tcn{n}) \rangle_{c_n}
    = x_{\nk,I}^{(n) (R,I)}\sigma_{y}^{(R,I)}( \eta_k^{c_n},c_{n-1})
    +\langle \yRI_{\nk} (\tcn{n}) \rangle_{c_{n-1}},
  \eeq
  \beq\label{esquemagenn2}
    \langle \pyRI (\tcn{n}) \rangle_{c_n} = x_{\nk,II}^{(n) (R,I)}
    \sigma_{\pi}^{(R,I)}(\eta_k^{c_n},c_{n-1})
    +\langle \pyRI_{\nk}(\tcn{n}) \rangle_{c_{n-1}}.
  \eeq

The second term of the r.h.s. of \eref{esquemagenn1} and \eref{esquemagenn2} is
the expectation value of the $(n-1)$-th collapse \textit{evaluated at
  the $n$-th collapse time } $\tcn{n}$. If we employ the matrix
notation introduced in the previous section, we can rewrite
\eref{esquemagenn1} and \eref{esquemagenn2} as
\begin{equation}\label{recetanmenos1}
  \Upsilon (\tcn{n},c_{n}) = \Delta(x_{\nk,i}^{(n) (R,I)}, \tcn{n},
  c_{n-1}) + \Upsilon (\tcn{n},c_{n-1}),
\end{equation}
where we introduced a new object
\begin{equation*}
  \Delta(x_{\nk,i}^{(n)
    (R,I)},\tcn{n}, c_{n-1}) \equiv \left( \begin{array}{c}

    x_{\nk,I}^{(n) (R,I)} \sigma_{y}^{(R,I)}( \eta_k^{c_n},c_{n-1}) \\
    x_{\nk,II}^{(n) (R,I)}\sigma_{\pi}^{(R,I)}( \eta_k^{c_n},c_{n-1})
  \end{array}
  \right),
\end{equation*}
with $i=I,II$.

\subsection{Evolution between collapses}

Equation \eref{matriz} characterizes the evolution of the state
between two successive collapses, e.g., $n$ and $n-1$.  In other
words, this means that \eref{matriz} is valid from $\tcn{n-1}$ (their
initial condition) to $\tcn{n}$.  We can rewrite \eref{matriz}, with
the notation adopted in this section, in order to see this more
clearly
\begin{equation}\label{matrizn}
  \Upsilon (\eta,c_n) = \textbf{U}(\eta,\tcn{n}) \Upsilon (\tcn{n},c_n).
\end{equation}
Thus, the \textit{evolution} from $\tcn{n-1}$ to $\tcn{n}$, of the
expectation values of the state $\ket{c_{n-1}}$, is determined by
\eref{matrizn} but using, as initial condition, the expectation value
given by the collapse $n-2$, which evolved in a similar manner. This
will lead us to a recursive relation for the dynamical equation of the
field's expectation value after $n$ collapses.  Note that this
description is just the orthodox quantum evolution following the
standard rules of quantum mechanics: between `measurements' the wave
function evolves following Schr\"odinger's equation, and at the times
when the `measurements' occur, the wave function is `collapsed' or
`reduced'. Then the wave function continues to evolve according to
Schr\"odinger's equation but now with the initial condition of the
`post-measurement' quantum state, etc.\

On account of the discussion above, we note that \eref{matrizn}
depends on the $(n-1)^{th}, (n-2)^{th},...,1^{st}$ collapse
states. Therefore, we will obtain a new expression for \eref{matrizn}
which will show this dependance explicitly.

We start by substituting \eref{recetanmenos1} in \eref{matrizn}
obtaining
\beq\label{colapson}
  \Upsilon (\eta,c_n) = \textbf{U}(\eta,\tcn{n}) \Delta(x_{\nk,i}^{(n)
    (R,I)},\tcn{n}, c_{n-1}) +\textbf{U}(\eta,\tcn{n}) \Upsilon
  (\tcn{n},c_{n-1}).
\eeq
The quantity $\Upsilon (\tcn{n},c_{n-1})$ contains information of the
expectation value of the fields in the state $\ket{c_{n-1}}$ at the
time $\tcn{n}$, but \eref{matrizn} give us the value of $\Upsilon$
for any time $\eta$ and any state $\ket{c_n}$. In other words, we can use
\eref{matrizn} and the collapse `recipe' \eref{recetanmenos1} ( to
obtain $\Upsilon(\tcn{n-1},c_{n-1})$) to calculate $\Upsilon
(\tcn{n},c_{n-1})$. This calculation will result in a term $\Upsilon
(\tcn{n-1},c_{n-2})$ which, again, can be computed from
\eref{matrizn} and \eref{recetanmenos1}, therefore, \eref{colapson}
is a recursive relation which depends explicitly from the very first
to the $(n-1)$-th post-collapse state. For example, if a single
collapse occurs, we have
\begin{equation}\label{colapso1}
  \Upsilon(\eta,c_1) = \textbf{U}(\eta,\tcn{1}) \Delta(x_{\nk,i}^{(1)
    (R,I)} , \tcn{1},c_0),
\end{equation}
because $|c_0 \rangle$ is taken to be the vacuum and
$\Upsilon(\tcn{1},c_0)=0$. For two collapses one obtains

\beq\label{colapso2}
  \Upsilon(\eta,c_2) = \textbf{U}(\eta,\tcn{2}) \Delta(x_{\nk,i}^{(2)
    (R,I)}, \tcn{2},c_1)  +\textbf{U}(\eta,\tcn{2}) \textbf{U}
  (\tcn{2},\tcn{1}) \Delta(x_{\nk,i}^{(1) (R,I)},\tcn{1},c_0).
\eeq

Thus, the general expression for $\Upsilon (\eta,c_n)$ after $n$
collapses is

  \bea\label{colapsonsimo} \Upsilon(\eta,c_n) &=
    \textbf{U}(\eta,\tcn{n}) \Delta(x_{\nk,i}^{(n)
      (R,I)},\tcn{n},c_{n-1}) \nonumber \\
      &+ \textbf{U} (\eta,\tcn{n})
    \textbf{U}(\tcn{n},\tcn{n-1}) \Delta
    (x_{\nk,i}^{(n-1) (R,I)},\tcn{n-1},c_{n-2}) \nonumber \\
   &+ \textbf{U} (\eta,\tcn{n}) \textbf{U}(\tcn{n},\tcn{n-1})
    \textbf{U} (\tcn{n-1},\tcn{n-2}) \Delta (x_{\nk,i}^{(n-2)
      (R,I)},\tcn{n-2},c_{n-3})+ \dots \nonumber \\
   &+ \textbf{U}(\eta,\tcn{n})\textbf{U} (\tcn{n},\tcn{n-1})
    \textbf{U} (\tcn{n-1},\tcn{n-2}) \textbf{U}(\tcn{n-2},\tcn{n-3})\times
    \dots \nonumber \\
    &\times \textbf{U} (\tcn{2},\tcn{1}) \Delta(x_{\nk,i}^{(1)
      (R,I)},\tcn{1},c_0).
  \eea

From \eref{matrizn} it is evident that the matrix $\textbf{U}
(\tcn{n},\tcn{n-1})$ represents the unitary evolution for the
expectation value of the fields in the state $\ket{c_{n-1}}$ from
$\tcn{n-1}$ to $\tcn{n}$. Because of the unitary evolution, we have
$\textbf{U}(\eta,\tcn{n})\textbf{U} (\tcn{n},\tcn{n-1}) \dots
\textbf{U} (\tcn{2},\tcn{1}) = \textbf{U}(\eta,\tcn{1})$. Using this
property in \eref{colapsonsimo}, we finally obtain
\begin{equation}\label{zncolapsos}
  \Upsilon(\eta,\tcn{n}) = \sum_{m=1}^n \textbf{U} (\eta,\tcn{m}) \Delta
  (x_{\nk,i}^{(m) (R,I)}, \tcn{m}, c_{m-1}).
\end{equation}

Equation \eref{zncolapsos} allows us to extract the evolution for the
expectation value of of $\pyRI_{\nk} (\eta) $ after $n$ collapses

  \bea
    \label{eq:valor_expectacion_pi_multicolapso}
    \langle \hat{\pi}_{\nk}^{(R,I)} (\eta) \rangle_{c_n} &=
    \sum_{m=1}^n \Bigg[-k\sin (k\eta-k\tcn{m}) x_{\nk,I}^{(m) (R,I)}
    \sigma^{(R,I)}_{y}(\tcn{m},c_{m-1}) \nonumber \\
    &+ \Bigg( \cos (k\eta-k\tcn{m}) + \frac{\sin
      (k\eta-k\tcn{m})}{k\tcn{m}} \Bigg) x_{\nk,II}^{(m) (R,I)}
    \sigma^{(R,I)}_\pi (\tcn{m},c_{m-1}) \Bigg].
  \eea

The result \eref{eq:valor_expectacion_pi_multicolapso} is the
generalization of \eref{eq:evolucion-vepi-2} for multiple collapses
during the inflationary epoch. We observe that the evolution of
$\pyRI_{\nk}$ resembles a superposition of many independent
one-collapse evolutions of the expectation values of $\pyRI_{\nk}$,
any of which suffered a collapse at different times.


As mentioned at the end of section \ref{observed}, to connect the
theoretical predictions with the observational quantities, we need to
compute $|\alpha_{lm}|^2_{M.L.}$ as given in \eref{alphalm}. That is,
we need to obtain $\overline{\expec{\hat \pi_{\nk} (\eta)}_{c_n}
  \expec{\hat \pi^{\dag}_{\nk'} (\eta)}_{c_n}}$, which will be our
next task.

First, we note that, in the notation introduced in
\eref{esquemagen1} and \eref{esquemagen2}, the expectation value of each
collapse scheme was decomposed generically as $x_{\nk,I}^{(1) (R,I)}
\sigma_{y}^{(R,I)}( \eta_k^{c_1}, c_0)$ and $x_{\nk,II}^{(1) (R,I)}
\sigma_{\pi}^{(R,I)}( \eta_k^{c_1}, c_0)$, where the random variables
are dimensionless, and $\sigma_y,\sigma_\pi$ are the part of the
collapse scheme carrying the units (e.g.,\ in the independent collapse
scheme $\sigma_{\pi}^{(R,I)}( \eta_k^{c_1}, c_0) = \sqrt{\hbar
  L^3/2}|g_k(\eta)|$ and $\sigma_{y}^{(R,I)}( \eta_k^{c_1}, c_0) =
\sqrt{\hbar L^3/2}|y_k(\eta)|$). The mean value of the product of
$x_{\nk,i}^{(n) (R,I)}$ depends on the particular scheme
considered. In the \emph{Independent} scheme, the product mean value
is given by
\beq\label{random}
\overline{x_{\nk,i}^{(n) R} x_{\nk',i}^{(m) R}} = (\delta_{\nk,\nk'} +
    \delta_{\nk,-\nk'})  \delta_{m,n}, \qquad \overline{x_{\nk,i}^{(m) I} x_{\nk',i}^{(n) I}} =
    (\delta_{\nk,\nk'} - \delta_{\nk,-\nk'}) \delta_{m,n},
\eeq
where $i=I,II$, and with all the other possible combinations equal zero.
Meanwhile, in the \emph{Newtonian} scheme it is
\beq\label{randomnew}
\overline{x_{\nk,II}^{(n) R} x_{\nk',II}^{(m) R}} = (\delta_{\nk,\nk'} +
    \delta_{\nk,-\nk'})  \delta_{m,n}, \qquad \overline{x_{\nk,II}^{(m) I}
      x_{\nk',II}^{(n) I}} = (\delta_{\nk,\nk'} - \delta_{\nk,-\nk'}) \delta_{m,n},
\eeq
and in the \emph{Wigner's} scheme we have
\beq\label{randomwig}
    \overline{x_{\nk}^{(n) R} x_{\nk'}^{(m) R}} = (\delta_{\nk,\nk'} +
    \delta_{\nk,-\nk'})  \delta_{m,n}, \qquad \overline{x_{\nk}^{(m) I}
      x_{\nk'}^{(n) I}} = (\delta_{\nk,\nk'} - \delta_{\nk,-\nk'}) \delta_{m,n}.
\eeq

The $\delta_{m,n}$ means that, in the three schemes, we are assuming
independency among the random variables associated to different
collapses respectively (e.g., the random variable $x_{\nk,I}^{(1) I}$
is independent of $x_{\nk,I}^{(2) I}$).


After choosing a particular collapse scheme (with the corresponding
characterization for the mean value of the random variables
\eref{random}, \eref{randomnew}, \eref{randomwig}), and recalling
that $\expec{\hat \pi_{\nk} (\eta)}_{c_n} = \expec{\hat \pi_{\nk}^{ R}
  (\eta)}_{c_n} + i \expec{ \hat \pi_{\nk}^{ I} (\eta)}_{c_n}$ (using
\eref{eq:valor_expectacion_pi_multicolapso}), one obtains the
quantity $\overline{\expec{\pk (\eta)}_{c_n}
  \expec{\hat{\pi}^\dag_{\nk'} (\eta)}_{c_n}}$ for each collapse
scheme.  The calculation will be simplified due to the fact that, as
usual, the average over the random variables (in the three collapse
schemes) will lead to a cancellation of the cross terms. Thus, after
going to the continuum limit ($L \rightarrow \infty$), the expression
for $|\alpha_{lm}|^2_{M.L.}$ \eref{alphalm}, after $N$ collapses is
given by
\begin{equation}\label{eq:alphalm generica multicolapso}
  |\alpha_{lm}|_{M.L.}^2 =
  \frac{4}{27 \pi} \frac{V \hbar^3}{\epsilon \mpl^4} \int \frac{dk}{k^2}
  |j_l (k R_D)|^2  \sum_{n=1}^NC^{(n)}_{l} (k,\eta_D),
\end{equation}
where $C^{(n)}_{l} (k,\eta_D)$ depends on the collapse scheme
considered. In the \emph{independent} scheme, this expression is

  \bea
    C^{(n)}_{l} (k,\eta_D) &= \bigg(k \sin (k\eta_D-k\tcn{n}) \bigg)^2
    \bigg( Y_{\nk}^+ + (-1)^{l} Y_{\nk}^- \bigg) \nonumber \\
    &+ \bigg(\cos (k\eta_D -k\tcn{n})
    + \frac{\sin (k\eta_D-k\tcn{n})}{k\tcn{n}} \bigg)^2 \bigg( \Pi_{\nk}^+ +
    (-1)^{l} \Pi_{\nk}^- \bigg),
  \eea

meanwhile, in the case of the \emph{Newtonian} scheme

\begin{equation}
  C^{(n)}_{l} (k,\eta_D) =  \bigg(\cos (k\eta_D -k\tcn{n})
  + \frac{\sin (k\eta_D-k\tcn{n})}{k\tcn{n}} \bigg)^2 \bigg( \Pi_{\nk}^+ +
  (-1)^{l} \Pi_{\nk}^- \bigg),
\end{equation}

the quantities $Y_{\nk}^{\pm}$ and $\Pi_{\nk}^{\pm}$ are defined as

  \begin{equation}
  \fl  Y_{\nk}^{\pm} \equiv (\Delta\yk^{R}(\eta_k^{c_n}))^2_{c_{n-1}} \pm
    (\Delta\yk^{I}(\eta_k^{c_n}))^2_{c_{n-1}}, \quad
    \Pi_{\nk}^{\pm} \equiv (\Delta\pk^{R}(\eta_k^{c_n}))^2_{c_{n-1}} \pm
    (\Delta\pk^{I}(\eta_k^{c_n}))^2_{c_{n-1}}.
  \end{equation}

Finally, in the \emph{Wigner's} scheme, $C^{(n)}_{l} (k,\eta)$ is
given by
\bea
  C^{(n)}_{l} (k,\eta_D) &= 2k^2 \Lambda_{k,n}^2 \bigg[ \sin (k\eta_D -
    k\tcn{n}) \sin \Theta_{k,n} \nonumber \\
     &+ \bigg( \cos (k\eta_D-k\tcn{n}) + \frac{\sin (k\eta_D-k\tcn{n})}{k\tcn{n}} \bigg) \cos \Theta_{k,n}
  \bigg]^2,
\eea

with $\Lambda_{k,n}$ and $\Theta_{k,n}$ defined as
\numparts
  \begin{equation}\label{eq:dispersion_tnc}
    \Lambda_{k,n}   \equiv  \frac{4\tcn{n} \sqrt{\hbar
        k}}{\sqrt{1+5(k\tcn{n})^2
        -\sqrt{1+10(k\tcn{n})^2+9(k\tcn{n})^4}}},
  \end{equation}
  \begin{equation}
    2  \Theta_{k,n}  \equiv
    \arctan\left(\frac{4k\tcn{n}}{1-3(k\tcn{n})^2}\right).
  \end{equation}
\endnumparts
It is worthwhile to comment that uncertainties in \eref{eq:alphalm
  generica multicolapso} are always evaluated at the $(n-1)$-th
collapse state.

Before discussing the physical implications of the general result
\eref{eq:alphalm generica multicolapso}, let us start by analyzing
the assumption of a single collapse in the \emph{independent} scheme,
in this case \eref{eq:alphalm generica multicolapso} reduces to

  \begin{equation}\label{eq:alphaN=1}
 \fl   |\alpha_{lm}|_{M.L.}^2 =
    \frac{4 V \hbar^3 }{54 \pi \epsilon \mpl^4} \int \frac{dk}{k} |j_l (k R_D)|^2
    \left( 1 + 2 \frac{ \sin^2 (k\eta_D-k\tcn{1}) }{(k \tcn{1})^2} +
      \frac{\sin 2 (k\eta_D-k\tcn{1})}{k\tcn{1}}\right).
  \end{equation}
  Considering again a single collapse and working within the
  \emph{Newtonian} scheme, \eref{eq:alphalm generica multicolapso}
  leads to
  \bea\label{eq:alphaN=1newt}
    |\alpha_{lm}|_{M.L.}^2 &=
    \frac{4 V \hbar^3 }{54 \pi \epsilon \mpl^4} \int \frac{dk}{k} |j_l (k
    R_D)|^2 \nonumber \\
    &\times \left[1 + \sin^2 (k\eta_D-k\tcn{1}) \left(
        \frac{1}{(k\tcn{1})^2} -1 \right) + \frac{\sin
        2(k\eta_D-k\tcn{1})}{k\tcn{1}} \right].
  \eea

Results \eref{eq:alphaN=1} and \eref{eq:alphaN=1newt} are consistent
with the findings presented in \cite{Sudarsky06a} and
\cite{Unanue2008}.  The result obtained from \eref{eq:alphalm
  generica multicolapso}, for a single collapse in the \emph{Wigner's}
scheme, also corresponds with the one presented in \cite{Unanue2008}.

We observe that, for the three schemes considered, in the case of a
single collapse $N=1$, only the uncertainties of the vacuum state
contribute to the integral in \eref{eq:alphalm generica
  multicolapso}.  The point is that for a single collapse,
\eref{eq:alphalm generica multicolapso} does not contain any
information characterizing the post-collapse state (the information
that defines a particular post-collapse state is contained in the
uncertainties evaluated in that precise state). That is, we do not
need to specify the post-collapse state. However, if we assume
multiple collapses, then the uncertainties of the post-collapse states
will contribute to the integral in \eref{eq:alphalm generica
  multicolapso}, and since the uncertainties will depend on the
pre-collapsed states, which are now different from the vacuum, then we
will need to specify every pre-collapse state (which will be the
subject of the next section).

An important feature arises in the \emph{independent} and
\emph{Newtonian} schemes, since for these cases, \eref{eq:alphalm
  generica multicolapso} exhibits an explicit dependence of $l$ (there is also another dependence on $l$ in the term
  $|j_l(kR_D)|^2$, however this dependence will not affect the
  compatibility of the theoretical predictions obtained in our
  approach with the ones from the standard treatment, since the latter, also involves this
  dependence on $l$ in the spherical Bessel function $j_l(kR_D)$) in the
terms
$Y_{\nk}^+ + (-1)^{l} Y_{\nk}^-$ and $\Pi_{\nk}^+ + (-1)^{l}
\Pi_{\nk}^-$. If $l$ is even, $Y_{\nk}^+ + (-1)^{l} Y_{\nk}^-=2
(\Delta\yk^{R}(\eta_k^{c_n}))^2_{c_{n-1}} $ and $\Pi_{\nk}^+ +
(-1)^{l} \Pi_{\nk}^-= 2 (\Delta\pk^{R}(\eta_k^{c_n}))^2_{c_{n-1}}$; if
$l$ is odd, $Y_{\nk}^+ + (-1)^{l} Y_{\nk}^-=2
(\Delta\yk^{I}(\eta_k^{cn}))^2_{c_{n-1}} $ and $\Pi_{\nk}^+ + (-1)^{l}
\Pi_{\nk}^-= 2 (\Delta\pk^{I}(\eta_k^{c_n}))^2_{c_{n-1}}$. Thus,
depending on the parity of $l$, the predicted quantity
$|\alpha_{lm}|_{M.L.}^2$ will involve the uncertainty of the real or
imaginary parts of $\hat{y}_\nk (\eta_k^{c_n})$ and $\hat{\pi}_\nk
(\eta_k^{c_n})$ which is not entirely compatible with the standard
prediction, namely a flat spectrum. In order to recover the standard
theoretical prediction, the dependence of $l$ should be avoided, and
the most natural option is that the uncertainties satisfy

  \begin{equation}\label{condinc}
  (\Delta\yk^{R}(\eta_k^{c_n}))^2_{c_{n-1}}=
    (\Delta\yk^{I}(\eta_k^{c_n}))^2_{c_{n-1}}, \qquad
    (\Delta\pk^{R}(\eta_k^{c_n}))^2_{c_{n-1}} =
    (\Delta\pk^{I}(\eta_k^{c_n}))^2_{c_{n-1}}.
  \end{equation}

It is clear, of course, that this is not the most generic
case\footnote{However, as we will show in the next section, if we
  assume that the post-collapse states are coherent states, this
  condition is fulfilled automatically.}, and needs not to be taken as
a necessary condition for the compatibility of the theoretical
predictions in our approach with the observations from the CMB, since
we still need to consider the physics of the cosmological epochs after
the end of the inflationary regime that leads to the so called
acoustic oscillations. It is also interesting to note, that the
condition on the uncertainties of the real and imaginary parts of
$\hat{y}_\nk (\eta_k^{c_n})$ and $\hat{\pi}_\nk (\eta_k^{c_n})$, only
applies to the \emph{independent} and \emph{Newtonian} schemes. In the
\emph{Wigner} scheme we do not find a similar condition for the
parameters $\Lambda_{k,n}$ and $\Theta_{k,n}$ that characterize the
uncertainties in that case.

Finally, we note that all the quantities involved in \eref{eq:alphalm
  generica multicolapso} are positive. In other words, we have a sum
of positive definite terms. Therefore, if we set $N \rightarrow
\infty$ the sum will generically diverge, which implies that we can
not set an infinite number of collapses because the predicted value
for $|\alpha_{lm}|^2_{M.L.}$ will tend to infinity. Thus we must restrict
consideration to the case with \textbf{ a finite number of
  collapses}. It is important to note that the calculations that lead
to result \eref{eq:alphalm generica multicolapso} have not considered
any particular post-collapse state. Of course (and we will do it in
the next section), we can consider a particular post-collapse state
and that information will enter in the uncertainties. However, the
conclusion obtained from \eref{eq:alphalm generica multicolapso}
related to the finiteness of the collapses is valid for a generic
post-collapse state in the three schemes considered.


\section{\label{caract}Characterization of the post-collapse states}

The information that characterizes a particular post-collapse state
will enter in the uncertainties of the field (and its momentum)
through the parameters that characterize the post-collapse
state. Thus, our first task will be to focus on obtaining the
uncertainties for the coherent and squeezed states and afterwards we
are going to use the results of the previous sections to obtain
predicted values for the observational quantities.

\subsection{\label{coherent}Coherent states as post-collapse states}

A simple election for a post-collapse state is a coherent state. A
coherent state is a specific state of the harmonic oscillator and its
dynamic is very similar to the one of the classic harmonic
oscillator. The coherent states $|\xi \rangle$ are defined as the
eigenstates of the annihilation operator $\ann$,
\begin{equation}
  \label{eq:coherent_def}
  \ann |\xi \rangle = \xi |\xi \rangle,
\end{equation}
since $\ann$ is not an Hermitian operator, $\xi$ is a complex number
and can be represented in complex polar form $\xi = |\xi|e^{i\chi}$,
where $|\xi|$ is the amplitude and $\chi$ is the phase.

Equation \eref{eq:coherent_def} physically implies that the coherent
state $\ket{\xi}$ is not affected by the detection and annihilation of
one particle. In a coherent state the quantum uncertainties of $\hat
p$ and $\hat q$ (the momentum and position of the quantum oscillator
respectively) take the minimum value, i.e.,\ $\Delta \hat p \Delta
\hat q = \frac{1}{2}\hbar$.

With the exception of the vacuum state $\ket{0}$ (which is also a
coherent state), every coherent state can be produced by the
application of the \emph{Displacement} operator
$\hat{D}(\xi)=\exp{(\xi\cre - \xi^*\ann)}$ to the vacuum state
\begin{equation*}
  |\xi \rangle=\hat D(\xi)\vac.
\end{equation*}

Using the simple properties of the coherent states we can calculate
the quantities $\dRI_{\nk}$, $\eRI_{\nk}$ and $\cRI_{\nk}$
\eref{eq:dce_def} when the post-collapse state of each mode of the
field is a coherent state $\ket{\xi_{\nk}}$,

\beq\label{eq: cde coherent}
\fl    d_{\nk, }^{(R,I)} = \xi_{\nk}^{(R,I)}, \qquad
     c_{\nk, }^{(R,I)} = (\xi_{\nk}^{(R,I)})^2, \qquad
      e_{\nk, }^{(R,I)} =|\xi_{\nk}^{(R,I)}|^2.
\eeq

Expressions \eref{eq: cde coherent}, \eref{eq:dy} and \eref{eq:dp} allow
us to obtain the evolution of the uncertainties of the field and its
conjugate momentum for any coherent state. Making use of the same
arguments that led \eref{matriz} to the generalization
\eref{matrizn} in the case of multiple collapses,
\eref{eq:dy} and \eref{eq:dp} can also be considered in conjunction with the
assumption that every post-collapse state is a coherent state
($\ket{\xi^{(n)}_{\nk}} = \ket{c_n}$)

    \bea\label{coherentes_n1}
\fl      (\Delta \hat y_{\nk}^{(R,I)}(\eta))^2_{c_n} &=& \Re [y_k^2(\eta)
      (\xi^{(n)(R,I)}_{\nk})^2] + \frac{1}{2} | y_k(\eta) |^2 (\hbar
      L^3 + 2
      |\xi^{(n) (R,I)}_{\nk}|^2) - 2 \Re[y_k(\eta) \xi^{(n)(R,I)}_{\nk}]^2 \nonumber \\
      &=& \frac{1}{2}|y_k(\eta)|^2 \hbar L^3 =\frac{\hbar L^3}{4k}
      \left( 1+\frac{1}{(k\eta)^2} \right) ,
    \eea
    \bea\label{coherentes_n2}
\fl      (\Delta \hat \pi_{\nk}^{(R,I)} (\eta))^2_{c_n} &=& \Re
      [g_k^2(\eta) (\xi^{(n)(R,I)}_{\nk})^2] + \frac{1}{2} | g_k(\eta)
      |^2 (\hbar L^3 + 2
      |\xi^{(n)(R,I)}_{\nk}|^2) - 2 \Re [g_k(\eta) \xi^{(n)(R,I)}_{\nk}]^2 \nonumber \\
      &=& \frac{1}{2} |g_k(\eta)|^2 \hbar L^3 = \frac{k \hbar L^3}{4}.
    \eea

This last result shows that the uncertainties of the $n$-th coherent
post-collapse state have the same form as those of the vacuum
state. We also note that the uncertainty of the conjugate momentum is
constant in the inflationary era.

\subsection{\label{squeezedsec}Squeezed states as post-collapse
  states}

Squeezed states can be considered as a more general case of the
coherent states. Qualitatively, a squeezed state is a state that has
the minimal uncertainty, not in the standard position and momentum
variables, but in a new pair of `rotated' canonical variables
(commonly referred as \emph{quadrature} variables
\cite{Walls1994}). Let us call them $\hat Q$ and $\hat P$. For a
squeezed state one can have `more (or less)' uncertainty in either
$\hat{Q}$ or $\hat{P}$, as long as their product is equal to the
minimum value allowed by Heisenberg's principle. The parameters of
the squeezed state control the angle of `rotation' and the `squeezing'
of the uncertainties.


The work with squeezed states is simplified by the introduction of the
following operator

\begin{equation}
  \hat{S}(\omega) \equiv \exp(\frac{1}{2}\omega^\star \hat{a}^2-\frac{1}{2}\omega\hat{a}^{\dag2}),
\end{equation}


where the parameter $\omega$ is a complex number. In particular,
$\omega$ can be written as $\omega = r e^{i \theta}$. The operator
$\hat{S}(\omega)$ is known as the \emph{Squeeze} Operator. Applying
the Squeeze and Displacement operators to the vacuum state we obtain a
squeezed state
\begin{equation}\label{c6}
  \ket{\xi \omega}  \equiv \hat{D}(\xi) \hat{S}
  (\omega) \vac.
\end{equation}
We note that the \emph{squeezed} state $\ket{\xi \omega }$ is
completely defined by four parameters: $|\xi|, \chi, r, \theta$

Some well known properties of the operators $\hat{D}(\xi)$ and
$\hat{S}(\omega)$ are

\begin{enumerate}
\item $\hat{D}^\dag(\xi)\hat{a}\hat{D}(\xi) = \hat{a} + {\xi}$.
\item $\hat{D}^\dag(\xi)\hat{a}^\dag\hat{D}(\xi) = \hat{a}^\dag +
  \overline{\xi}$.
\item $\hat{S}^\dag(\omega)\hat{a}\hat{S}(\omega) = \hat{a}\cosh{r} -
  \hat{a}^{\dag}e^{-i\theta}\sinh{r}$.
\item $\hat{S}^\dag(\omega)\hat{a}^\dag\hat{S}(\omega) =
  \hat{a}^\dag\cosh{r} - \hat{a}e^{i\theta}\sinh{r}$.
\item Both $\hat{D}$ and $\hat{S}$ are unitary operators.
\end{enumerate}

By regarding the post-collapse state of each mode as a \emph{squeezed}
state and using the properties 1-5, one can obtain $d_{\nk}^{(R,I)}$,
$c_{\nk}^{(R,I)}$ and $e_{\nk}^{(R,I)}$ from \eref{eq:dce_def}

  \begin{equation}\label{eq:d edo squeeze}
    d_{\nk}^{(R,I)} = \xi_{\nk}^{(R,I)},
  \end{equation}
  \beq\label{eq:c edo squeeze}
    c_{\nk}^{(R,I)}= -\hbar
    L^3\cosh{r_{\nk}^{(R,I)}}\sinh{r_{\nk}^{(R,I)}}e^{-i\theta_{\nk}^{(R,I)}}+
    (\xi_{\nk}^{(R,I)})^2,
  \eeq
  \beq\label{eq:e edo squeeze}
    e_{\nk}^{(R,I)} =\hbar L^3 \sinh^2{r_{\nk}^{(R,I)}}+|\xi_{\nk}^{(R,I)}|^2.
  \eeq

Equations \eref{eq:dy} and \eref{eq:dp} give us the evolution of the
uncertainties, in terms of the quantities $d_{\nk}^{(R,I)}$,
$c_{\nk}^{(R,I)}$ and $e_{\nk}^{(R,I)}$. As in the coherent state,
\eref{eq:dy} and \eref{eq:dp} can be generalized straightforward to the case
of multiple collapse. Thus, by considering the post-collapse states of
each mode as \emph{squeezed} states, we can substitute \eref{eq:d edo squeeze}, \eref{eq:c edo squeeze} and \eref{eq:e edo squeeze}   into \eref{eq:dy} and \eref{eq:dp}, which for multiple
collapses yields

    \bea\label{eq:dysqueezed}
      (\dyk^{(R,I)} (\eta))^2_{c_n} &= \frac{\hbar L^3}{4k} \bigg[ 1 +
      \frac{1}{(k\eta)^2} \bigg] \bigg\{- \sinh (2r^{c_n (R,I)}_{\nk}) \nonumber \\
      &\times \cos \left[\theta^{c_n (R,I)}_{\nk} + 2 \arctan
          \left(\frac{1}{k\eta}\right)+ 2k\eta \right] + \cosh (2r^{c_n
          (R,I)}_{\nk}) \bigg\},
    \eea
    \begin{equation}\label{eq:dpsqueezed}
      (\dpk^{(R,I)} (\eta))^2_{c_n} = \frac{\hbar L^3 k}{4} \bigg[ \sinh
      (2r^{c_n (R,I)}_{\nk}) \cos ( \theta^{c_n (R,I)}_{\nk} + 2k\eta) +
      \cosh (2r^{c_n (R,I)}_{\nk}) \bigg].
    \end{equation}

The squeezing parameters $r^{c_n (R,I)}_{\nk}$ and $\theta^{c_n
  (R,I)}_{\nk}$ refer to the squeeze parameters of the $n$-th
post-collapse squeezed state of each mode. The situation at hand is
totally different from the coherent case, in which the uncertainties
are completely characterized by the vacuum state despite $n$ collapses
have occurred. In the squeeze state case, it is evident from
\eref{eq:dysqueezed} and \eref{eq:dpsqueezed} that the dispersions
$(\Delta\hat{y}_{\nk}^{(R,I)}(\eta))^2_{c_n},(\Delta\hat{\pi}_{\nk}^{(R,I)}
(\eta))^2_{c_n}$ are determined by the squeeze parameters
$r_{\nk}^{c_n (R,I)}$ and $\theta_{\nk}^{c_n (R,I)}$. This is a
crucial difference with the coherent case in which the uncertainties
are independent of the parameters characterizing the coherent
state.

\subsection{Connections with the observational quantities}

The uncertainties of the field are characterized by both, the
particular post-collapse state and the collapse scheme. In the rest of
this section we will focus on the \emph{independent} collapse scheme,
however, similar conclusions as those obtained from these results
can be derived when considering the other two collapse schemes that
have been proposed so far.

\subsubsection{\label{obssqueezed}Squeezed States as postcollapse states}
The connection with the observations will be made under the following
assumptions: I) The wave-function of the field has collapsed $N$ times
and the $N$ post-collapse states are squeezed states. II) The
uncertainties $(\Delta\yk^{R}(\eta_k^{c_n}))^2_{c_{n-1}}$ and
$(\Delta\yk^{I}(\eta_k^{c_n}))^2_{c_{n-1}}$ are equal (as well as the
uncertainties $(\Delta\pk^{R}(\eta_k^{c_n}))^2_{c_{n-1}}$ and
$(\Delta\pk^{I}(\eta_k^{c_n}))^2_{c_{n-1}}$), which is motivated by
the discussion at the end of section \ref{multiples}.

Under the assumption II), \eref{eq:alphalm generica multicolapso}
(recall that we are working under the \emph{independent} scheme) takes
the simplified form

  \bea
    \label{eq:alphalm generica dos colapsos}
    |\alpha_{lm}|_{M.L.}^2 &= \frac{8}{27 \pi} \frac{V\hbar^3
    }{\epsilon \mpl^4} \int \frac{dk}{k^2} |j_l (k R_D)|^2  \sum_{n=1}^N\bigg[ \bigg( k \sin (k\eta_D-k\tcn{n}) \bigg)^2
    (\Delta\yk
    (\eta_k^{c_n}))^2_{c_{n-1}} \nonumber \\
    &+ \bigg(\cos (k\eta_D-k\tcn{n}) + \frac{\sin (k\eta_D-k\tcn{n})
    }{k\tcn{n}} \bigg)^2 (\Delta\pk (\eta_k^{c_n}))^2_{c_{n-1}}
    \bigg].
  \eea
  After a little algebra, the expression for $|\alpha_{lm}|_{M.L.}^2$
  obtained by substituting \eref{eq:dysqueezed} and \eref{eq:dpsqueezed} in \eref{eq:alphalm generica
    dos colapsos} becomes
  \bea
    \label{eq:alphalm squeeze n colapsos}
     \fl  |\alpha_{lm}|_{M.L.}^2 &=
  \frac{2}{27 \pi} \frac{V\hbar^3 }{\epsilon \mpl^4} \int
    \frac{dk}{k} |j_l (k R_D)|^2
  \sum_{n=1}^N \Bigg[\bigg( 1 + \frac{\sin
      2(k\eta_D-k\tcn{n}) }{k\tcn{n}} \bigg) \nonumber \\
    &\times   \bigg( \cosh
    2r^{c_{n-1}}_{\nk} + \sinh 2r^{c_{n-1}}_{\nk} \cos
    (\theta^{c_{n-1}}_{\nk} + 2k\tcn{n})
    \bigg)
    + \frac{2 \sin^2 (k\eta_D-k\tcn{n}) }{(k\tcn{n})^2} \nonumber \\
    &\times \bigg( \cosh
    2r^{c_{n-1}}_{\nk} + k\tcn{n} \sinh 2r^{c_{n-1}}_{\nk} \sin
    (\theta^{c_{n-1}}_{\nk} + 2k\tcn{n}) \bigg ) \Bigg].
  \eea

The above result lead us to conclude that, in order to obtain a
reasonable power spectrum, that is, a nearly flat Harrison-Zel'dovich
spectrum, there seems to be one simple case characterized by two
particular conditions:

First, for each one of the $n$ post-collapse states, $k\tcn{n}$ should
be independent of $k$ but dependent of $n$. I.e., the time of collapse
for the $n$ post-collapse states of the different modes should depend
on the mode's frequency according to $\eta^{c_n}_{k} = f_n/k$ (where $f_n$ is a real number that changes for each collapse). This
condition is the generalization for $n$ collapses of the result presented in \cite{Sudarsky06a} where a
single collapse was considered (a possible deviation of such `recipe'
for the time of collapse was studied in \cite{Unanue2008}). In other words, the result
\eref{eq:alphalm squeeze n colapsos} generalizes the condition
$\eta^c_{k} \propto 1/k$ in the case of multiple collapses.

Second, a nearly flat spectrum is recovered if the parameters
characterizing the $n$ squeezed states are also independent of the
mode's frequency, that is, if $r^{c_n}_{\nk}=r^{c_n}$ and
$\theta^{c_n}_{\nk}=\theta^{c_n}$ are independent of $\nk$ but
dependent of $n$. This does not mean that the uncertainties for each
mode are all the same, because the uncertainties are also
characterized by the time of collapse of each mode and its frequency,
as can be seen in \eref{eq:dysqueezed} and \eref{eq:dpsqueezed}.

\subsubsection{An upper bound for the number of collapse using
  coherent states as post-collapse states}

As already noted generically, the number of collapses in each mode
must be finite, and we expect to provide a simple estimate in this
subsection. We will continue the consideration of the
\emph{independent} collapse scheme, but we will assume that all the
$N$ post-collapse states are \emph{coherent} states (which, after all,
are just a particular class of squeezed states with $r_{\nk}=0$). That
is, we can use the uncertainties \eref{coherentes_n1} and \eref{coherentes_n2} to obtain a
predicted value for $|\alpha_{lm}|_{M.L.}^2$. Note, however, that in
the case of coherent states, assumption (II) of the subsection
\ref{obssqueezed} is naturally obtained because the uncertainties of
every coherent state are equal to the uncertainties of the vacuum
state which automatically satisfy $(\Delta\yk^{R}(\eta_k))^2_{c_0} =
(\Delta\yk^{I}(\eta_k))^2_{c_0}$ (as well as
$(\Delta\pk^{R}(\eta_k))^2_{c_0} =
(\Delta\pk^{I}(\eta_k))^2_{c_0}$). Substituting \eref{coherentes_n1} and \eref{coherentes_n2}
in \eref{eq:alphalm generica multicolapso} yields

  \bea\label{eq:alphalm_coherent}
\fl    |\alpha_{lm}|_{M.L.}^2 =  \frac{2}{27 \pi} \frac{V\hbar^3 }{\epsilon
      \mpl^4} \int \frac{dk}{k} |j_l (k R_D)|^2 \sum_{n=1}^N\bigg( 1
    + \frac{\sin 2(k\eta_D-k\tcn{n})  }{k\tcn{n}}  +  \frac{2 \sin^2
      (k\eta_D-k\tcn{n}) }{(k\tcn{n})^2} \bigg). \nonumber \\
  \eea

From this last expression is a relatively simple task to obtain
information regarding the maximum number of collapses allowed by
observations. If we assume that $|k\tcn{n}| \gg k\eta_D$, that is,
the time for the $1^{st},2^{nd},...,N^{th}$ collapse occurs at
very early stage of the inflationary regime;
\eref{eq:alphalm_coherent} is approximated by
\begin{equation}\label{eq:alphalm_coherent2}
  |\alpha_{lm}|_{M.L.}^2 \approx
  \frac{2}{27 \pi} \frac{V\hbar^3}{\epsilon \mpl^4}  \int \frac{dk}{k} |j_l (k
  R_D)|^2N,
\end{equation}
using that $\int x^{-1} j^2_l(x) dx = \pi/l(l+1)$, the expression
above reduces to
\begin{equation}\label{eq:alphalm_coherent3}
  |\alpha_{lm}|_{M.L.}^2 \approx   \frac{2}{27 } \frac{V
    \hbar^3}{\epsilon \mpl^4}  \frac{N}{l(l+1)}.
\end{equation}

In general $|\alpha_{lm}|_{M.L.}^2 $ is independent of $m$ and the
quantity that is presented as the result of observations is $OB_{l} =
l(l+1)C_l$, where $C_l = (2l+1)^{-1} \sum_m |\alpha_{lm}^{obs}|^2$. If
we ignore the physics of the plasma that follows after the reheating
era, $OB_l$ is essentially independent of $l$ corresponding to the
amplitude of the metric perturbations (which is roughly
$10^{-10}$). Thus, setting $OB_l \equiv A$, the maximum number of
collapses $N_{max}$ allowed by the observations is
\begin{equation}
  N_{max} \approx \frac{27 \epsilon \mpl^4 A}{2V\hbar^3 }.
\end{equation}

We believe that this constraint might be of great help in studying the
viability of the actual proposals for the detail physical mechanism
that lies behind the collapse we have been considering.

\section{\label{discusion}Discussion}

As first reviewed in \cite{Sudarsky06a}, the inflationary account of
the origin of cosmic structure posses a serious shortcoming, namely,
the emergence of structure from an initial state that was homogeneous
and isotropic. The proposal to address this existing issue was through
the introduction of a modification of standard quantum theory
corresponding to a dynamical reduction of the wave function. The
present study represents a continuation of the investigation of such
proposal.

In this paper, we have examined the possibility that multiple
collapses take place in each of the modes of the quantum field.  This
study required a much more detailed characterization of the
post-collapse states. This, in turn, required the introduction of
extra assumptions. We focused here in the possibility that the states
are coherent or squeezed and under these assumptions we were able to
further constrain, beyond the results of previous analyses, the
features of the collapse hypothesis required for agreement with
observations.  These we will discuss in the following.

The first result obtained in this manuscript is that in order to
recover a flat spectrum, and assuming that multiple collapses occur,
then the uncertainties of the real and imaginary parts of the
fluctuation of the inflaton field, i.e., $\hat{y}_\nk$ and its
conjugated momentum $\hat{\pi}_\nk$, must be equal. We can interpret
this result as the most natural option for selecting simple candidates
for post-collapse states since the uncertainty of each mode of the
field and its conjugated momentum is characterized by specifying the
post-collapse state. Therefore, given a particular state $|\Xi
\rangle$ for each mode $\nk$, one can calculate the uncertainties of
the field and its conjugated momentum for that state. If the
uncertainties for each mode satisfy the relation
$(\Delta\yk^{R})^2_{\Xi}=(\Delta\yk^{I})^2_{\Xi}$ (as well as
$(\Delta\pk^{R})^2_{\Xi} = (\Delta\pk^{I})^2_{\Xi})$ , then $|\Xi
\rangle$ can be regarded as a reasonable choice for a post-collapse
state. In fact, in section \ref{coherent}, we found that, for coherent
states, the relation between the uncertainties of the real and
imaginary parts of $\hat{y}_\nk$ and $\hat{\pi}_\nk$ is satisfied
automatically. Consequently, a coherent state is a natural candidate
for a post-collapse state. The fact that a coherent state acts as a
good candidate for a post-collapse state is consistent with the notion
that a coherent state of the field is the closest quantum mechanical
state to a classical description of the field, i.e., a state for which
the semiclassical approximation of gravity given by $G_{ab} = 8\pi G
\langle \hat{T}_{ab} \rangle$ is valid in the sense of Ehrenfest's
theorem and thus qualifies for a reasonable candidate for a
post-collapse state.

Nevertheless, for a generic squeezed state $|\Sigma \rangle$,
$(\Delta\yk^{R})^2_{\Sigma} \neq (\Delta\yk^{I})^2_{\Sigma}$ and
$(\Delta\pk^{R})^2_{\Sigma} \neq (\Delta\pk^{I})^2_{\Sigma}$, but this
does not mean that post-collapse squeezed states are forbidden. That
is, one can select a set of squeezed states, characterized by the
squeezing parameters $r_{\nk}^{(R,I)}$ and $\theta_{\nk}^{(R,I)}$,
such that $r_{\nk}^{R}=r_{\nk}^{I}$ and $\theta_{\nk}^{R} =
\theta_{\nk}^{I}$ for which the relation in the uncertainties
holds. Furthermore, in section \ref{obssqueezed} we argued that, given a
collection of multiple post-collapse squeezed states characterized by
$r_{\nk}^{c_n}$ and $\theta_{\nk}^{c_n}$, then, the simplest choice
that allows the recovering of the standard flat spectrum, is that the
squeezing parameters be independent of $\nk$. The point is that we
again used the observations as a guide to uncover the particular
characteristics of a squeezed state that could be regarded as a
reasonable post-collapse state.  We should note that, as discussed
in \cite{Unanue2008}, we can not expect such a strict pattern to be
followed in an exact manner in a theory involving a collapse
controlled by some fundamentally random events, and as such one can in
principle investigate the effects of the expected deviations on the
observational data. The investigation of the detail signature of those
deviations, as well as the observational bounds on them
(i.e. analogues of those considered in \cite{Unanue2008}), is part of
our ongoing research program.


Another important result from this work is that the number of
collapses must be finite under generic conditions. However we could,
in principle, select a set of post-collapse states and adjust the
uncertainties of the field (and its conjugated momentum) and the times
of collapse in a way that the predicted observational quantity (the
sum in \eref{eq:alphalm generica multicolapso}) would remain finite,
even for an infinite number of collapses. Evidently, this would amount
to a fine tune of the scheme which we do not see as an attractive
choice.  On the other hand, we should say that if the collapse of the
state, which gives birth to the inhomogeneities observed in the CMB,
is a process that keeps occurring indefinitely even after inflation
ends, the Newtonian potentials would also be changing, thus affecting
in a rather random way the propagation of photons from the last
scattering surface to our satellites.  These ideas might be considered
as related, at least at the phenomenological level, to those explored
in \cite{Dowker2010}.  We did not investigate these issues here.  In
the present work we rather concentrated on the generic sort of
conditions for the collapse during inflation and found, not only that
the number of collapses should be finite, but obtained, -- under the
extra hypothesis on the form of post-collapse states--
a rough estimate on the number of collapses in terms of the parameters
of the inflaton potential.

All of the previous discussion shows that, even though in principle we
do not know precisely what is the nature of the physics behind what we
call the collapse, we can, in fact, obtain some insights on the
`rules' that govern it, i.e., those determining the nature of
post-collapse states and the number of collapses of each mode, by
comparing the observations with our theoretical predictions.

We are beginning to investigate the possible connection of our
proposal with other more developed collapse mechanisms involving
similar non unitary modifications of quantum theory. Henceforth, the
path to follow in our future research is to explore the connections of
our proposal with other collapse mechanisms compatible with the
conclusions obtained in this and previous works.

We believe that, in the case of the inflationary paradigm, we cannot
content ourselves with the fact that calculations lead to results that
match the observations but which can not be fully justified within the
context of the interpretations provided by our current physical
theories. We readily acknowledge that, although our proposal seems to
offer a clearer picture of the emergence of the seeds of cosmic
structure, it might be ultimately an incorrect proposal which might
need to be replaced by something even more complex and distant from
the established physical paradigms.  What seems clear is that the
standard account of the genesis of the cosmic structure, something
intimately tied with the rise in the conditions that are a
prerequisite for our own existence, is not fully satisfactory and that
on the other hand, our present and future access to detailed empirical
data makes the issue not only susceptible to scientific inquire, but
from our point of view, one of the most promising fertile grounds
where some fundamental questions can be explored.

\ack
 GL and DS acknowledge support from PAPPIT project IN 119808 and
  CONACyT project No 101712.  DS was supported in part by sabbatical
  fellowships from CONACyT and DGAPA-UNAM and the hospitality of the
  IAFE. ADU acknowledges support from Redes temáticas de investigación CONACyT `Red Complejidad, Ciencia y Sociedad'.

\section*{References}
\bibliography{colapso}

\providecommand{\newblock}{}
\begin{thebibliography}{10}
\expandafter\ifx\csname url\endcsname\relax
  \def\url#1{{\tt #1}}\fi
\expandafter\ifx\csname urlprefix\endcsname\relax\def\urlprefix{URL }\fi
\providecommand{\eprint}[2][]{\url{#2}}

\bibitem{Padmanabhan96}
Padmanabhan T 1996 {\em Cosmology and Astrophysics - Through Problems\/}
  (Cambridge University Press)

\bibitem{Mukhanov2005}
Mukhanov V 2005 {\em Physical Foundations of Cosmology\/} (New York: Cambridge
  University Press)

\bibitem{Weinberg2008}
Weinberg S 2008 {\em Cosmology\/} (New York: Oxford University Press)

\bibitem{Liddle2009}
Lyth D and AR L 2009 {\em The Primordial Density Perturbation: Cosmology,
  Inflation and the Origin of Structure\/} (New York: Cambridge University
  Press)

\bibitem{Kiefer98a}
Kiefer C and Joos E 1998 Decoherence: Concepts and examples (\textit{Preprint}
  \eprint{quant-ph/9803052v1})
  \urlprefix\url{http://www.citebase.org/abstract?id=oai:arXiv.org:quant-ph/98%
03052}

\bibitem{Kiefer98b}
Kiefer C, Lesgourgues J, Polarski D and Starobinsky A~A 1998 The coherence of
  primordial fluctuations produced during inflation (\textit{Preprint}
  \eprint{gr-qc/9806066v1})
  \urlprefix\url{http://www.citebase.org/abstract?id=oai:arXiv.org:gr-qc/98060%
66}

\bibitem{Kiefer00a}
Kiefer C 2000 {\em Nucl.Phys.Proc.Suppl\/} {\bf 88} 255--258 (\textit{Preprint}
  \eprint{astro-ph/0006252})

\bibitem{Kiefer08}
Kiefer C and Polarski D 2008 Why do cosmological perturbations look classical
  to us? (\textit{Preprint} \eprint{0810.0087v1})
  \urlprefix\url{http://www.citebase.org/abstract?id=oai:arXiv.org:0810.0087}

\bibitem{Sudarsky09}
Sudarsky D 2009 {Shortcomings in the Understanding of Why Cosmological
  Perturbations Look Classical} (\textit{Preprint} \eprint{0906.0315})

\bibitem{Sudarsky06a}
Perez A, Sahlmann H and Sudarsky D 2006 {\em Class. Quantum Grav.\/} {\bf 23}
  2317--2354 (\textit{Preprint} \eprint{gr-qc/0508100})

\bibitem{Sudarsky06b}
Sudarsky D 2007 {\em J. Phys.: Conf. Ser.\/} {\bf 68} (\textit{Preprint}
  \eprint{gr-qc/0612005})

\bibitem{Sudarsky07a}
Sudarsky D 2007 {\em PoSQG-Ph:038\/} (\textit{Preprint}
  \eprint{arXiv:0712.2795})

\bibitem{Sudarsky07}
Sudarsky D 2007 {\em J. Phys.: Conf. Ser.\/} {\bf 67} (\textit{Preprint}
  \eprint{gr-qc/0701071v1})

\bibitem{Unanue2008}
De~Un\'anue A and Sudarsky D 2008 {\em Phys. Rev. D\/} {\bf 78}
  (\textit{Preprint} \eprint{0801.4702})

\bibitem{Unanue2010}
{De Un\'anue} A 2010 {\em The quantum origins of the cosmological asymmetry\/}
  Ph.D. thesis Universidad Nacional Aut\'onoma de M\'exico (\textit{Preprint}
  \eprint{1010.5765})

\bibitem{Gabriel2010}
Le\'on G and Sudarsky D 2010 {\em Class. Quantum Grav.\/} {\bf 27} 225017
  (\textit{Preprint} \eprint{1003.5950v2})

\bibitem{Diosi1984}
Di{\'{o}}si L 1984 {\em Phys. Lett.\/} {\bf 105A} 199--202

\bibitem{Diosi}
Di{\'{o}}si L 1989 {\em Phys. Rev A\/} {\bf 40} 1165--1174

\bibitem{DiosiLajos07}
Di{\'{o}}si L 2007 {\em Journal of Physics A Mathematical and Theoretical\/}
  {\bf 40} 2989 (\textit{Preprint} \eprint{quant-ph/0607110})

\bibitem{Penrose96}
Penrose R 1996 {\em Gen. Rel. Grav.\/} {\bf 28} 581--600

\bibitem{Penrose94}
Penrose R 1994 {\em Shadows of the Mind. A search for the missing science of
  consciousness\/} (Oxford University Press)

\bibitem{Penrose02}
Penrose R 1989 {\em The Emperors New Mind\/} (Oxford University Press)

\bibitem{Penrose05}
Penrose R 2004 {\em {The Road to Reality. A complete guide to the laws of the
  universe}\/} (Londres: Jonathan Cape)

\bibitem{Ghirardi1986}
Ghirardi G, Rimini A and Weber T 1986 {\em Physical Review D\/} {\bf 34} 470

\bibitem{Ghirardi1990}
Ghirardi G 1990 {\em Physical Review A\/} {\bf 42} 1057

\bibitem{Bassi02}
Bassi A and Ghirardi G 2002 {\em Physical Review A\/} {\bf 65} 042114
  (\textit{Preprint} \eprint{quant-ph/0201122v1})
  \urlprefix\url{http://www.citebase.org/abstract?id=oai:arXiv.org:quant-ph/02%
01122}

\bibitem{Bassi03}
Bassi A and Ghirardi G 2003 {\em Physics Reports\/} {\bf 379} 257
  (\textit{Preprint} \eprint{quant-ph/0302164v2})
  \urlprefix\url{http://www.citebase.org/abstract?id=oai:arXiv.org:quant-ph/03%
02164}

\bibitem{Bassi07-2}
Bassi A, Ghirardi G~C and Salvetti D~G~M 2007 {\em Mathematical Systems
  Theory\/} {\bf 40} 13755 (\textit{Preprint} \eprint{0707.2940v3})
  \urlprefix\url{http://www.citebase.org/abstract?id=oai:arXiv.org:0707.2940}

\bibitem{Nakamura08}
Nakamura K 2008 {\em Phys.Rev.D\/} {\bf 80} 124021 (\textit{Preprint}
  \eprint{0804.3840v4})
  \urlprefix\url{http://www.citebase.org/abstract?id=oai:arXiv.org:0804.3840}

\bibitem{Birrel94}
Birrel N and Davies P 1994 {\em Quantum fields in curved space\/} (Cambridge:
  Cambridge University Press)

\bibitem{Bardeen83}
Bardeen J~M, Steinhardt P~J and Turner M~S 1983 {\em Phys. Rev.\/} {\bf D28}
  679

\bibitem{Brandenberger84}
Brandenberger R~H 1985 {\em Rev. Mod. Phys.\/} {\bf 57} 1

\bibitem{Lyth85}
Lyth D~H 1985 {\em Phys. Rev.\/} {\bf D31} 1792--1798

\bibitem{Mukhanov90}
Mukhanov V~F, Feldman H~A and Brandenberger R~H 1992 {\em Phys. Rept.\/} {\bf
  215} 203--333

\bibitem{Walls1994}
Walls D and Milburn G 1994 {\em Quantum Optics\/} (Springer-Verlag)

\bibitem{Dowker2010}
Contaldi R~C, Dowker F and Philpott L 2010 {\em Class. Quantum Grav.\/} {\bf
  27} 172001 (\textit{Preprint} \eprint{1001.4545})

\end{thebibliography}
\bibliographystyle{iopart-num}

\end{document}